\documentclass[a4paper,12pt]{article}
\usepackage{amsmath}
\usepackage{amssymb}
\usepackage{amsthm}
\usepackage{latexsym}
\usepackage{mathrsfs}
\usepackage[dvips]{graphicx}
\usepackage{color}
\usepackage{subfigure}
\usepackage{hyperref}

\def\f(#1){{\mathop{f}^{(#1)}}}
\def\m(#1){{\mathop{m}^{(#1)}}}
\def\C(#1){{\mathop{C}^{(#1)}}}
\def\p(#1){{\mathop{p}^{(#1)}}}

\def\ben{\begin{equation}}
\def\een{\end{equation}}
\def\bena{\begin{eqnarray}}
\def\eena{\end{eqnarray}}
\def\non{\nonumber}

\def\S{{\bf S}}

\def\C{{\cal C}}

\def\D{D}

\def\mr{{\mathbb R}}
\def\A{{\bf A}}

\def\T{{\bf T}}
\newcommand{\V}{{\cal Y}}
\newcommand{\myid}{{\bf 1}}
\newcommand{\ran}{{\rm ran}\,}
\renewcommand{\ker}{{\rm ker}\,}
\newcommand{\mn}{{\mathbb N}}

\newcommand{\mc}{{\mathbb C}}

\newcommand{\id}{id}

\renewcommand{\H}{{\mathcal H}}
\newcommand{\F}{{\mathcal F}}
\renewcommand{\a}{{\mathbf b}}

\renewcommand{\D}{{\mathcal D}}

\newcommand{\e}{{\rm e}}

\theoremstyle{definition}

\newtheorem{thm}{Theorem}

\newtheorem{lemma}{Lemma}

\newtheorem{prop}{Proposition}

\newtheorem{defn}{Definition}[section]

\begin{document}

\title{Quantum field theory in terms of consistency conditions I:\\
General framework, and perturbation theory via Hochschild cohomology}

\author{Stefan Hollands\thanks{HollandsS@Cardiff.ac.uk}\\
School of Mathematics\\
Cardiff University, UK
}

\maketitle

\begin{abstract}
In this paper, we propose a new framework for quantum field theory
in terms of consistency conditions. The consistency conditions that
we consider are ``associativity'' or ``factorization'' conditions
on the operator product expansion (OPE) of the theory, and are proposed to be
the defining property of any quantum field theory. Our framework
is presented in the Euclidean setting, and is applicable in principle
to any quantum field theory, including non-conformal ones. In our framework,
we obtain a characterization of perturbations of a given quantum
field theory in terms of a certain cohomology ring of Hochschild-type.
We illustrate our framework by the free field, but our constructions
are general and apply also to interacting quantum field theories.
For such theories, we propose a new scheme to construct the OPE which 
is based on the  use of non-linear quantized field equations.
\end{abstract}

\pagebreak

\section{Introduction}

Quantum field theory has been formulated in different ways, the most
popular ones being the path-integral approach and the operator formalism.
In the path integral approach, one aims to construct the correlation
functions of the theory as the moments of some measure on the space of classical
field configurations. In the operator formalism, the quantum fields are
viewed as linear operators which can act on physical states.

The path integral has the advantage of being closely related to classical field theory.
In fact, the path integral measure is, at least formally, directly given in terms
of the classical action of the theory. The operator formalism is more useful in
contexts where no corresponding classical theory---and hence no Lagrange formalism---is
known for the quantum field theory. It has been used extensively in the context
of conformal or integrable field theories in two spacetime dimensions. In the
operator formalism, one may take the point of view that the theory is determined
by the algebraic relations between the quantum field observables. This viewpoint was
originally proposed in a very abstract form by Haag and Kastler, see e.g.~\cite{Haag}. Other
proposals aimed in particular at conformal field theories include e.g. the
approach via vertex operator algebras due to Borcherds, Frenkel, Lopowski, Meurman
and others~\cite{vertex1,vertex2,vertex3,vertex4}, see also a related proposal by Gaberdiel and Goddard~\cite{Gaberdiel}.
A different approach of an essentially algebraic nature applicable
to "globally conformally invariant quantum field theories" in $D$ dimensions is due to~\cite{Rehren1,Rehren2}.
Approaches emphasizing the algebraic relations between the fields have also
turned out to fundamental to the construction of quantum field theories on general curved backgrounds~\cite{HW01,HW02,BF00,BFV03}, because in this case there is no preferred Hilbert
space representation or vacuum state.

One way to encode the algebraic relations between the fields in a very
explicit way (at least at short
distances) is the Wilson operator product expansion (OPE)~\cite{Wilson,WZ,Zimmermann}.
This expansion is at the basis of the modern treatments of two-dimensional conformal
field theory, and it is a key tool in the quantitative analysis of asymptotically free
quantum gauge theories in four dimensions such as Quantum Chromo Dynamics. The OPE  can also be established
for perturbative quantum field theory in general curved spacetimes~\cite{Hollands06}.
In this reference,  it was observed in particular that the OPE coefficients satisfy
certain "asymptotic clustering" or "factorization" relations when various groups of points in the operator products are
scaled together at different rates. This observation was taken one step further in~\cite{HW08},
where it was suggested that the OPE should in fact be viewed as the
fundamental datum describing a quantum field theory on curved (and flat) spacetimes, and that the factorization
conditions should be viewed as the essential constraints upon the OPE coefficients.

In this paper, we will analyze these constraints on the OPE coefficients, and thereby
formulate a new approach to quantum field theory in terms of the resulting consistency conditions.
One of our main new points is that all these constraints can be encoded in a single
condition which is to be viewed as an analogue of the
usual "associativity condition" in ordinary algebra. We then show that
it is possible to give a new formulation of perturbation theory which directly
involves the OPE coefficients, but does not directly use such notions---and is more general as---path
integrals or interaction Lagrangians. This new approach relies on a perturbative formulation of the
consistency condition and is hence essentially algebraic in nature.
Its mathematical framework is a certain cohomology of "Hochschild type"
which we will also set up in this paper. If our approach to perturbation theory
is combined with the assumptions of certain linear or non-linear field equations,
then a constructive algorithm is obtained to determine the terms in the perturbation
series order-by-order. We expect that our approach is equivalent to more standard
ones despite its rather different appearance, but we do not investigate this
issue in the present paper.

Some of our ideas bear a (relatively remote) resemblance to
ideas that have been proposed a long time ago within the
``bootstrap-approach'' to conformally invariant quantum field
theories, where constraints of a somewhat similar, but not identical,
nature as ours have been considered under the name ``crossing
relations"~\cite{Migdal, Polyakov, Todorov, Mack1, Mack2}.
But we stress from the outset that our approach is aimed at all quantum field theories---including
even quantum field theories on generic spacetime manifolds without symmetries---and not just conformal ones as in
these references. The ideas on the use of non-linear field equations expressed in section~\ref{interactingfields}
also bear a resemblance to a constructive method in quantum field theory introduced by Steinmann (see e.g.~\cite{steinmann}), but
he is mainly concerned with the Wightman functions rather than the OPE, which is a key difference. Some
of the ideas in section~\ref{interactingfields} were developed, in preliminary form, in extensive discussions
with N.~Nikolov during his tenure as a Humboldt fellow at the U. of G\" ottingen in 2005/2006, see also
the notes~\cite{Nikolovnotes}. In the present form described in section~\ref{interactingfields}, these
ideas were developed in collaboration with H.~Olbermann, and more details will be given in 
a future paper~\cite{Olbermann}.

This paper is organized as follows. We first explain in sec.~2 the
basic ideas of this paper, namely, the idea of that the
factorization conditions may be expressed by a single associativity
condition, the new formulation of perturbation theory in
our framework, the generalization to gauge field theories, and
the approach via field equations. These ideas are then
explained in detail in the subsequent sections.

\section{Basic ideas of the paper}

The operator product expansion states that the product of two operators
may be expanded as
\ben\label{basic}
\phi_{a}(x_1) \phi_{b}(x_2) = \sum_c C_{ab}^c(x_1, x_2) \, \phi_c(x_2) \, ,
\een
where $a, b, c$ are labels of the various composite quantum fields
$\phi_{a}$ in the theory. This relation is intended to
be valid after taking expectation values in any (reasonable) state
in the quantum field theory. The states, as well as the OPE coefficients
typically have certain analytic continuation properties that
arise from the spectrum condition in the quantum field theory. These
properties imply that the spacetime arguments may be continued to
a real Euclidean section of complexified Minkowski spacetime, and
we assume this has been done. An important condition on
the OPE coefficients arises when one
considers the operator product expansion of $3$ operators (in the Euclidean domain),
\ben\label{3ops}
\phi_{a} (x_1) \phi_{b}(x_2) \phi_{c}(x_3)  = \sum_d
C_{abc}^d(x_1, x_2, x_3) \, \phi_d(x_3) \, .
\een
Let us consider a situation where one pair of points is closer to each other than another pair of points.
For example, let $r_{23}$ be
smaller than $r_{13}$, where
\ben
r_{ij} = |x_i - x_j|
\een
is the Euclidean distance between point $x_i$ and point $x_j$. Then we expect that we can
first expand the operator product $\phi_{b}(x_2) \phi_{c}(x_3)$ in eq.~\eqref{3ops}
around $x_3$, then multiply by $\phi_{a}(x_1)$, and finally expand the resulting product
around $x_3$. We thereby expect to obtain the relation
\ben\label{2ops1}
C_{abc}^d(x_1, x_2, x_3) = \sum_e C_{bc}^e(x_2, x_3) C_{ae}^d(x_1, x_3)
\een
Similarly, if
$r_{12}$ is smaller than $r_{23}$, we expect that we
can first expand the operator product $\phi_{a}(x_1) \phi_{b}(x_2)$ around $x_2$,
then multiply the result by $\phi_{c}(x_3)$, and finally expand again around $x_3$.
In this way, we expect to obtain the relation
\ben\label{2ops2}
C_{abc}^d(x_1, x_2, x_3) = \sum_e C_{ab}^e(x_1, x_2) C_{ec}^d(x_2, x_3) \, .
\een
A consistency relation now arises because on the open domain $r_{12} < r_{23} < r_{13}$
both expansions~\eqref{2ops1}, \eqref{2ops2} must be valid and therefore should
coincide. Thus, we must have
\ben\label{assoccomp}
\sum_e C_{ab}^{e}(x_1, x_2) C_{e c}^{d}(x_2, x_3) =
\sum_e C_{bc}^{e}(x_2, x_3) C_{a e}^{d}(x_1, x_3) \,
\een
when $r_{12} < r_{23} < r_{13}$.
This requirement imposes a very stringent condition on the
OPE-coefficients. We will refer to this condition as a "consistency-" or "associativity" condition. The basic idea of this paper is that this condition on the 2-point OPE coefficients
incorporates the full information about the structure of the quantum field theory.
Therefore, conversely, if a solution to the consistency condition
can be found, then one has in effect constructed a quantum field theory.
We will pursue this idea below in the following different directions.

\subsection{Coherence}

First, we will pursue the question whether any further consistency conditions
in addition to eq.~\eqref{assoccomp} can arise when one considers products
of more than three fields, by analogy with the analysis just given for three fields. For
example, if we consider the OPE of four fields $\phi_{a}(x_1) \phi_{b}(x_2)
\phi_{c}(x_3) \phi_{d}(x_4)$ and investigate the possible different
subsequent expansions of such a product in a similar manner as above, we
will get new relations for the 2-point OPE coefficients analogous to eq.~\eqref{assoccomp}.
These will now involve four points and correspondingly more factors of the 2-point
OPE coefficients. Are these conditions genuinely new, or do they already follow from the
relation~\eqref{assoccomp}?

As we will argue, this question is analogous to the
question whether, in an ordinary algebra, there are new constraints on the product
coming from "higher order associativity conditions". As in this analogous
situation, we will see that in fact no new
conditions arise, i.e. the associativity condition~\eqref{assoccomp} is
the only consistency condition. We will also see that all higher order expansion
coefficients such as $C_{abcd}^e(x_1, x_2, x_3, x_4)$ are
uniquely determined by the 2-point OPE coefficients. Thus, in this sense,
the entire information about the quantum field theory is contained in
these 2-point coefficients $C^c_{ab}(x_1, x_2)$, and the entire
set of consistency conditions is coherently encoded in the associativity
condition~\eqref{assoccomp}.

For this reason, we call the result
a "coherence theorem", by analogy with the well-known similar result
in algebra and in category theory~\cite{MacLane}.
These results are described in detail in sec.~\ref{coherence}.

\subsection{Perturbation theory as Hochschild cohomology}\label{subsecpert}

Given that the 2-point OPE coefficients $C^c_{ab}(x_1, x_2)$ are
 considered in as the fundamental entities in quantum field theory in
 our approach, it is interesting
to ask how to formulate perturbation theory in terms of these coefficients.
For this, we imagine that we are given a 1-parameter family of these coefficients
parametrized by $\lambda$. For each $\lambda$, the coefficients should
satisfy the associativity condition~\eqref{assoccomp}, and for $\lambda=0$,
the coefficients describe the quantum field theory that we wish to perturb around.
We now expand the 1-parameter family of OPE-coefficients in a Taylor- or perturbation series
in $\lambda$, and we ask what constraints the consistency condition will impose upon the
Taylor coefficients. In order to have a reasonably uncluttered notation, let us use an
"index free" notation for the OPE-coefficients suppressing the indices $a,b,c,\dots$.
Thus, let us view the 2-point OPE coefficients $C^c_{ab}(x_1, x_2)$ as the
components of a linear map $\C(x_1, x_2): V \otimes V \to V$,
where $V$ is the vector space whose basis elements are in one-to-one correspondence
with the composite fields $\phi_a$ of the theory. The Taylor expansion is
\ben\label{Cexpansioncomp}
\C(x_1, x_2; \lambda) = \sum_{i=0}^\infty
\C_i(x_1, x_2) \, \lambda^i \, .
\een
We similarly expand the associativity condition as a power series in $\lambda$.
If we assume that the associativity condition is fulfilled at zeroth order, then
the corresponding condition for the first order perturbation of the
2-point OPE-coefficients is given by
\bena\label{firstocons}
&&\C_0(x_2, x_3)\Big(\C_1(x_1, x_2) \otimes \id \Big) - \C_0(x_1, x_3)\Big(
\id \otimes \C_1(x_2, x_3) \Big) + \non\\
&&\C_1(x_2, x_3)\Big(\C_0(x_1, x_2) \otimes \id \Big) - \C_1(x_1, x_3)\Big(
\id \otimes \C_0(x_2, x_3) \Big) = 0
\, ,
\eena
for $r_{12} < r_{23} < r_{13}$, in an obvious tensor product notation.
As we will see, this condition is of a cohomological nature,
and the set of all first order perturbations satisfying this condition modulo
trivial perturbations due to field redefinitions
can be identified with the elements of a certain cohomology ring which we will
define in close analogy to Hochschild cohomology~\cite{MacLane1,Happel,Connes}.
Similarly, the conditions for the higher order perturbations can also be
described in terms of this cohomology ring. More precisely, at each order
there is a potential obstruction to continue the perturbation series---i.e.,
to satisfy the associativity condition at that order---and this
obstruction is again an element of our cohomology ring.

In practice, $\lambda$ can be e.g. a parameter that measures the strength of
the self interaction of a theory, as in the theory characterized by
the classical Lagrangian $L = (\partial \varphi)^2 + \lambda \varphi^4$. In this example,
one is perturbing around a free field theory, for which the OPE-coefficients are
known completely. Another example is when one perturbs around a more general
conformal field theory---not necessarily described by a Lagrangian. Yet another
example is when $\lambda = 1/N$, where $N$ is the number of "colors" of a
theory, like in $SU(N)$~Yang-Mills theory. In this example, the theory that one is perturbing
around is the large-$N$-limit of the theory.

These constructions are described in detail in sec.~\ref{perturbations}.

\subsection{Local gauge theories}

Some modifications must be applied to our constructions when one is dealing with
theories having local gauge invariance, such as Yang-Mills theories.
When dealing with such theories, one typically has to proceed in two steps. The first step is
to introduce an auxiliary theory including further fields. For example, in pure
Yang-Mills theory, the auxiliary theory has as basic fields the 1-form gauge
potential $A$, a pair of anti-commuting "ghost fields" $U, \bar U$, as well as another
auxiliary field $F$, all of which take values in a Lie-algebra.
Having constructed the auxiliary theory, one then removes the
additional degrees of freedom in a second step, thereby arriving at the
actual quantum field theory one is interested in. The necessity of
such a two-step procedure can be seen from many viewpoints, maybe most directly
in the path-integral formulation of QFT~\cite{Faddeev}, but also actually even
from the point of view of classical Hamiltonian field theory, see e.g.~\cite{Henneaux}.

As is well-known, a particularly elegant and useful way to implement this two-step procedure
is via the so-called BRST-formalism~\cite{Becci}, and this is also the most
useful way to proceed in our approach to quantum field theory via the OPE.
In this approach one defines, on the space of auxiliary fields,
a linear map $s$ ("BRST-transformation"). The crucial properties of this map
are that it is a symmetry of the auxiliary theory, and that it is nilpotent,
$s^2 = 0$. In the case of Yang-Mills theory it is given by
\ben\label{BRSTt}
sA = dU - i\lambda[A, U]\, , \quad sU = -\frac{i\lambda}{2}[U,U] \, , \quad
s\bar U = F \, , \quad sF=0 \, ,
\een
on the basic fields and extended to all monomials in the basic fields and
their derivatives ("composite fields") in such a way that $s^2 = 0$. In our formalism,
the key property of the auxiliary theory is now that the map $s$ be compatible
with the OPE of the auxiliary theory.
The condition that we need is that, for any product of composite fields, we have
\ben
s[\phi_{a}(x_1) \phi_{b}(x_2)] = [s \phi_{a}(x_1)] \phi_{b}(x_2) \pm
\phi_{a}(x_1) s\phi_{b}(x_2) \, ,
\een
where the choice of $\pm$ depends on the Bose/Fermi character of the fields under
consideration. If we apply the OPE to the products in this equation, then it
translates into a compatibility condition between the OPE coefficients $C_{ab}^c(x_1, x_2)$ and the
map $s$. This is the key condition on the auxiliary theory beyond the associativity
condition~\eqref{assoccomp}. As we show, it enables one to pass
from the auxiliary quantum field theory to true gauge theory by taking a
certain quotient of the space of fields.

We will also perform a perturbation analysis of gauge theories. Here,
one needs not only to expand the OPE-coefficients [see eq.~\eqref{Cexpansioncomp}],
but also the BRST-transformation map $s(\lambda)$, as perturbations will typically change the form
of the BRST transformations as well---seen explicitly for Yang-Mills
theory in eqs.~\eqref{BRSTt}. We must now satisfy at each order in perturbation
theory an associativity condition as described above,
and in addition a condition which ensures compatibility of the perturbed BRST map and the
perturbed OPE coefficients at the given order.
As we will see, these conditions can again be
encoded elegantly and compactly in a cohomological framework.

These ideas will be explained in detail in sec.~\ref{hochschild}.

\subsection{Field equations}\label{subfieldeq}

The discussion so far has been focussed so far on the general mathematical structures behind
the operator product expansion. However, it is clearly also of interest to construct the
OPE coefficients for concrete theories. One way to describe a theory is via classical field
equations such as
\ben\label{phicubed}
\square \varphi = \lambda \varphi^3 \, ,
\een
where $\lambda$ is a coupling parameter. One may exploit such relations
by turning them into conditions on the OPE coefficients. The OPE coefficients are then determined
by a ``bootstrap''-type approach. The conditions implied by eq.~\eqref{phicubed} arise as follows:
We first view the above field equation as a relation between quantum fields, and we multiply by an
arbitrary quantum field $\phi_a$ from the right:
\ben
\square \varphi(x_1) \phi_a(x_2) = \lambda \varphi^3(x_1) \phi_a(x_2) \, .
\een
Next, we perform an OPE of the expressions on both sides, leading to the relation
$\square C_{\varphi a}^b = \lambda C_{\varphi^3  a}^b$. As explained above in subsection~\ref{subsecpert},
each OPE coefficient itself is a formal power series in $\lambda$, so this equation clearly
yields a relationship between different orders in this power series. The basic idea is to exploit
these relations and to derive an iterative construction scheme.

To indicate how this works, it is useful to introduce, for each field $\phi_a$,
a ``vertex operator'' $\V(x, \phi_a)$, which is a linear map on the space $V$ of
all composite fields. The matrix components of this linear map are simply given
by the OPE coefficients, $[\V(x, \phi_a)]_b^c = C_{ab}^c(x,0)$, for details see
sec.~\ref{leftrep}. Clearly, the vertex
operator contains exactly the same information as the OPE coefficient. In the above theory,
it is a power series $\V = \sum \V_i \lambda^i$ in the coupling. The field equation then
leads to the relation
\ben
\square \V_{i+1}(x, \varphi) = \V_{i}(x,\varphi^3) \, .
\een
The zeroth order $\V_0$ corresponds to the free theory, described in sec.~\ref{freefield},
and the higher order ones are determined inductively by inverting the Laplace operator.
To make the scheme work, it is necessary to construct $\V_{i}(x, \varphi^3)$ from
$\V_{i}(x,\varphi)$ at each order. It is at this point that we need the consistency
condition. In terms of the vertex operators, it implies e.g. relations like
\ben
\sum_{j=0}^i \V_j(x, \varphi) \V_{i-j}(y, \varphi) = \sum_{j=0}^i \V_j(y, \V_{i-j}(x-y, \varphi)\varphi) \, .
\een
On the right side, we now use a relation like $\V_0(x-y, \varphi)\varphi = \varphi^2 + \dots$.
Such a relation enables one to solve for $\V_i(y, \varphi^2)$ in terms of inductively
known quantities. Iterating this type of argument, one also obtains $\V_i(y, \varphi^3)$, and
in fact any other vertex operator at $i$-th order. In this way, the induction loop closes.

Thus, we obtain an inductive scheme from
the field equation in combination with the consistency condition.
 At each order, one has to perform one---essentially trivial---inversion
of the Laplace operator, and several infinite sums implicit in the consistency condition. 
These sums arise when composing two vertex operators if these are written in terms of their 
matrix components. Thus, to
compute the OPE coefficients at
$n$-th order in perturbation theory, the "computational cost" is roughly
to perform $n$ infinite sums. This is
similar to the case of ordinary perturbation theory, where at $n$-th order one
has to perform a number of Feynman integrals increasing with $n$. Note however that,
by contrast with the usual approaches to perturbation theory, our procedure
is completely well-defined at each step. Thus,
there is no "renormalization" in our approach in the sense of "infinite counterterms".

The details of this new approach to perturbation theory are outlined in sec.~\ref{interactingfields}, and
presented in more detail in a forthcoming paper with H. Olbermann.

\pagebreak

\section{Axioms for quantum field theory}\label{axiomatic}

Having stated the basic ideas in this paper in an informal
way, we now turn to the precise formulation of these ideas. For this, we begin
in this section by explaining our axiomatic setup for quantum field theory.
The setup we present here is broadly speaking the same as that
presented in~\cite{HW08}. In particular, the key idea here as well as
in~\cite{HW08} is that the operator product expansion (OPE)
should be regarded as the defining property of a quantum field theory.
However, there are some differences to~\cite{HW08} in that
we work on flat space here (as opposed to a general curved spacetime), and
we also work in a Euclidean framework. As a consequence, the microlocal conditions
stated in~\cite{HW08} will be replaced by analyticity conditions, the commutativity
condition will be replaced by a symmetry condition and
the associativity conditions on the OPE coefficients
will be replaced by conditions on the existence of various
power series expansions.

The first ingredient in our definition of a quantum field theory is an infinite-dimensional
vector space, $V$. The elements in this vector space are to be thought of as the
components of the various composite scalar, spinor, and tensor fields in the theory.
For example, in a theory describing a single real scalar field $\varphi$, the elements of
$V$ would be in one-to-one correspondence with the monomials of $\varphi$ and its
derivatives [see sec.~\ref{freefield}].
The space $V$ is assumed to be graded in various ways which reflect the possibility to classify
the different composite quantum fields in the theory by their spin, dimension,
Bose/Fermi character, etc. First, for Euclidean quantum field theory on $\mr^D$,
the space $V$ should carry a representation of the rotation group $SO(D)$ in
$D$ dimensions respectively of its covering group ${\rm Spin}(D)$ if
spinor fields are present. This representation should decompose into
unitary, finite-dimensional irreducible representations (irrep's) $V_{S}$,
which in turn are characterized
by  the corresponding eigenvalues $S=(\lambda_1, \dots, \lambda_r)$ of the $r$
Casimir operators associated with $SO(D)$. For $D=2$, this
is a weight $w \in \mr$, for $D=3$ this is an integer or half-integer
spin, and for $D=4$ this is a pair of spins (using the isomorphism between
$SU(2) \times SU(2)$ and the covering of the 4-dimensional rotation group).
Thus we assume that $V$ is a graded vector space
\ben\label{decomp}
V = \bigoplus_{\Delta \in \mr_+} \bigoplus_{S \in {\rm irrep}}
\mc^{N(\Delta, S)} \otimes V_{S} \, .
\een
The numbers $\Delta \in \mr_+$ provide an additional grading which
will later be related to the "dimension" of the quantum fields. The natural
number $N(\Delta, S)$ is the multiplicity of the quantum fields
with a given dimension $\Delta$ and spins $S$. We assume this multiplicity
to be finite. As always in this paper, the infinite sums in this decomposition
are understood without any closure taken, i.e., the elements of $V$ are
in one-to-one correspondence with sequences of the form
$(|v_1 \rangle, |v_2\rangle, \dots, |v_n\rangle, 0, 0, \dots)$ with only finitely many non-zero entries,
where $|v_i\rangle$ is a vector in the $i$-th summand in the decomposition.
On the vector space $V$, we would like to have an anti-linear, involutive
operation called $\star: V \to V$ which should be thought of as taking the hermitian adjoint of
the quantum fields. We would also like to have a linear grading map $\gamma: V \to V$
with the property $\gamma^2 = id$. The vectors corresponding to eigenvalue $+1$ are
to be thought of as "bosonic", while those corresponding to eigenvalue $-1$ are to
be thought of as "fermionic".

So far, we have only defined a list of objects---in fact a linear space---that we think
of as labeling the various composite
quantum fields of the theory. The dynamical content and quantum nature of
the given theory is now incorporated in the operator product coefficients associated with
the quantum fields. This is a hierarchy denoted
\ben
\C = \bigg( \C(x_1, x_2), \C(x_1, x_2, x_3), \C(x_1, x_2, x_3, x_4), \dots \bigg) \, ,
\een
where each $\C(x_1, \dots, x_n)$ is an analytic function on the "configuration space"
\ben
M_n := \{(x_1, \dots, x_n) \in (\mr^D)^n \mid x_i \neq x_j \quad
\text{for all $1 \le i< j \le n$}\} \, ,
\een
taking values in the linear maps\footnote{Strictly speaking, $\C(x_1, \dots, x_n)$
does not take its values in the space $V$, because for each $v_1, \dots, v_n \in V$,
the expression $C(x_1, \dots, x_n)(v_1 \otimes \dots \otimes v_n)$ typically has non-zero
components in an infinite number of summands in the decomposition~\eqref{decomp}.
By contrast, $V$ by definition only consists of vectors which have non-zero components
only for finitely many summands. Thus, $\C(x_1, \dots, x_n)$ actually takes
values in the larger space ${\rm Hom}(V^*, \mc) \supset V$, where $V^*$ is the
(algebraic) dual of $V$, see eq.~\eqref{dualdecomp}.
}
\ben
\C(x_1, \dots, x_n) : V \otimes \cdots \otimes V \to V \, ,
\een
where there are $n$ tensor factors of $V$. For one point, we set $\C(x_1) = id: V \to V$,
where $id$ is the identity map.
The components of these maps in a basis of $V$ correspond to the OPE coefficients
mentioned in the previous section.
More explicitly, if $\{
|v_a \rangle \}$ denotes a basis of $V$ adapted to the grading of $V$,
and $\{\langle v^a |\}$ the corresponding basis of the dual space
\ben\label{dualdecomp}
V^* = \bigoplus_{\Delta \in \mr_+} \bigoplus_{S \in {\rm irrep}} \mc^{N(\Delta, S)}
\otimes V_{\overline S} \, ,
\een
with $V_{\overline S}$ denoting the conjugate representation,
$\langle v^b| v_a \rangle = \delta^b_a$, then
\ben\label{Ccompdef}
C^b_{a_1 \dots a_n}(x_1, \dots, x_n) = \langle v^b| \C(x_1, \dots, x_n)|
v_{a_1} \otimes \dots \otimes v_{a_n} \rangle \,\,\,\,,
\een
using the standard bra-ket notations such as $|v_{a_1} \otimes \dots \otimes v_{a_n}\rangle
:= |v_{a_1} \rangle \otimes \cdots \otimes |v_{a_n} \rangle$.
The basic properties of quantum field theory are now expressed as the following properties
on the OPE coefficients:

\medskip
\noindent
\paragraph{\bf Hermitian conjugation:} Denoting by $\iota: V \to V$ the
anti-linear map given by the star operation $\star$, we have
\ben
\overline{\C(x_1, \dots, x_n)} = \iota \, \C(x_1, \dots, x_n) \, \iota^n
\een
where $\iota^n := \iota \otimes \cdots \otimes \iota$ is the $n$-fold tensor
product of the map $\iota$.

\medskip
\noindent
\paragraph{\bf Euclidean invariance:} Let $R$ be the representation
of ${\rm Spin}(D)$ on $V$,
let $a \in \mr^D$ and let $g \in {\rm Spin}(D)$. Then we have
\ben
\C(gx_1 + a, \dots, gx_n + a) = R^*(g)
\, \C(x_1, \dots, x_n) \, R(g)^n \, ,
\een
where $R(g)^n$ stands for the $n$-fold tensor product $R(g) \otimes \dots \otimes R(g)$.

\medskip
\noindent
\paragraph{\bf Bosonic nature:} The OPE-coefficients should themselves be
"bosonic" in the sense that
\ben
\C(x_1, \dots, x_n) = \gamma \, \C(x_1, \dots, x_n) \, \gamma^n
\een
where $\gamma^n$ is again a shorthand for the $n$-fold tensor product
$\gamma \otimes \dots \otimes \gamma$.

\medskip
\noindent
\paragraph{\bf Identity element:} There exists a unique element
$\myid$ of $V$ of dimension $\Delta = 0$,
with the properties $\myid^\star = \myid, \gamma(\myid) = \myid$,
such that
\ben\label{iidop}
\C(x_1, \dots, x_n)|v_1 \otimes \cdots \myid \otimes \cdots v_{n-1} \rangle =
\C(x_1, \dots \widehat{x_i}, \dots x_n) |v_1 \otimes \cdots \otimes v_{n-1}\rangle \, .
\een
where $\myid$ is in the $i$-th tensor position, with $i \le n-1$. When $\myid$
is in the $n$-th tensor position, the analogous formula takes a slightly
more complicated form. This is because $x_n$ is the point around which we expand
the operator product, and therefore this point and the corresponding $n$-th
tensor entry is on a different footing than the other points and tensor entries.
To motivate heuristically the appropriate form of the identity axiom in this case, we start by
noting that, if $\phi_a$ is a quantum (or classical) field, then we can formally
perform a Taylor expansion
\ben\label{taylor}
\phi_a(x_1) = \sum_{i=0}^\infty \frac{1}{i!} y^{\mu_1} \cdots y^{\mu_i}
\partial_{\mu_1} \dots \partial_{\mu_i} \phi_a(x_2) \, ,
\een
where $y=x_1-x_2$.
Now, each field $\partial_{\mu_1} \dots \partial_{\mu_i} \phi_a$ is
just another quantum field in the theory---denoted, say by $\phi_b$ for some label $b$---so trivially,
we might write this relation alternatively in the form
$\phi_a(x_1) = \sum t_a^b(x_1, x_2) \phi_b(x_2)$. Here, $t_a^b$ are defined by the above
Taylor expansion, up to potential trivial
changes in order to take into account the fact that in the chosen labeling of the fields,
a derivative of the field $\phi_a$
might actually correspond to a linear combination of other fields. Now formally, we have
\bena
\sum_b C^b_{a_1 \dots a_{n-1} \myid}(x_1, \dots, x_n) \, \phi_b(x_n) &=&
\phi_{a_1}(x_1) \cdots \phi_{a_{n-1}}(x_{n-1}) \myid \\
&=& \sum_b C^b_{a_1 \dots a_{n-1}}(x_1, \dots, x_{n-1}) \, \phi_b(x_{n-1}) \non\\
&=& \sum_{c,b} C^c_{a_1 \dots a_{n-1}}(x_1, \dots, x_{n-1}) \, t_c^b(x_{n-1}, x_n) \, \phi_b(x_{n}) \,, \non
\eena
so we are led to conclude that
\ben\label{unitcomp}
C^b_{a_1 \dots a_{n-1} \myid}(x_1, \dots, x_n) = \sum_c t^b_c(x_{n-1}, x_n) \,
C^c_{a_1 \dots a_{n-1}}(x_1, \dots, x_{n-1}) \, .
\een
Note that, in eq.~\eqref{taylor}, the operators on the right contain derivatives and
are thus expected to have a dimension that is not smaller than that of the operator
on the right hand side. It thus follows that $t^a_b(x_1, x_2)$ can only be nonzero
if the dimension of the operator $\phi_a$ is not less than the dimension of $\phi_b$.
Since there are only finitely many operators up to a given dimension, it follows that
the sum in eq.~\eqref{unitcomp} is finite, and there are no convergence issues.

We now abstract the features that we have heuristically derived. We postulate
the existence of a "Taylor expansion map", i.e. a
linear map\footnote{Here, the same remarks apply as in the
 footnote 1.} $t(x_1, x_2): V \to V$ for each $x_1, x_2 \in \mr^D$ with the
following properties. The map should transform in the same
way as the OPE coefficients, see the Euclidean invariance axiom. If $V^\Delta$ denotes the
subspace of $V$ in the decomposition~\eqref{decomp} spanned by vectors
of dimension $\Delta$, then
\ben
t(x_1, x_2) V^\Delta \subset \bigoplus_{\widehat \Delta \ge \Delta} V^{\widehat \Delta} \, .
\een
Furthermore, we have the cocycle relation
\ben
t(x_1, x_2) t(x_2, x_3) = t(x_1, x_3) \, .
\een
The restriction of any vector of $t(x_1, x_2) V^\Delta$ to any subspace $V^{\widehat \Delta}$
should have a polynomial dependence on $x_1-x_2$.
Finally, for each $v_1, \dots, v_{n-1} \in V$, we have
\ben
\C(x_1, \dots, x_n)|v_1 \otimes \dots v_{n-1} \otimes \myid \rangle =
t(x_{n-1}, x_n) \C(x_1, \dots, x_{n-1}) |v_1 \otimes \dots \otimes v_{n-1}\rangle \, ,
\een
for all $(x_1, \dots, x_n) \in M_n$. This is the desired formulation
for the identity axiom when the identity operator is in the $n$-th position.
Note that this relation implies in particular the relation
\ben
t(x_1, x_2) |v\rangle = \C(x_1, x_2)| v \otimes \myid \rangle  \, ,
\een
i.e., $t(x_1, x_2)$ uniquely determines the 2-point OPE coefficients with
an identity operator and vice-versa. In particular, we have $t(x_1, x_2) \myid = \myid$
using the eq.~\eqref{iidop} and $\C(x_1) = id$, meaning that the identity operator does not
depend on a "reference point".

\medskip
\noindent
\paragraph{\bf Factorization:} Let $I_1, \dots, I_r$ be a partition of the
set $\{1, \dots, n\}$ into disjoint ordered subsets, with the property
that all elements in $I_i$ are greater than all elements in $I_{i-1}$ for
all $i$. For example, for $n=5$, such a partition is
$I_1 = \{1\}, I_2 = \{2,3,4\}, I_3 = \{5,6\}$.
For each ordered subset $I \subset \{1, \dots, n\}$, let
$X_I$ be the ordered tuple $(x_i)_{i \in I} \in (\mr^D)^{|I|}$,
let $m_k = {\rm max}(I_k)$, and set $\C(X_I) := id$ if
$I$ is a set consisting of only one element. Then we have
\ben\label{factorization}
\C(X_{\{1, \dots, n\}}) = \C(X_{\{m_1,\dots,m_r\}}) \Big(
\C(X_{I_1}) \otimes \cdots \otimes \C(X_{I_r})
\Big)
\een
as an identity on the open domain
\bena\label{domaindef}
\D[\{I_1, \dots, I_r\}] &:=& \bigg\{
(x_1, \dots, x_n) \in M_n \mid \non\\
&& {\rm min} \, d(X_{\{m_1, \dots,m_r\}}) > {\rm max}\, (d(X_{I_1}), \dots, d(X_{I_r}))
\bigg\} \, .
\eena
Here, $d(X_I)$ denotes the set of relative distances between points
of points in a collection $X_I = (x_i)_{i \in I}$, defined as
the collection of positive real numbers
\ben
d(X_I) := \{ r_{ij} \,\, \mid i,j \in I, i \neq j \} \, .
\een
Note that the factorization identity~\eqref{factorization} when expressed in
a basis of $V \otimes \dots \otimes V$ involves an $r$-fold infinite sum on the
right side. The factorization property is in particular the statement that these
infinite sums converge on the indicated domain. No statement is made about the
convergence outside the domain, and in fact the series are expected
to diverge outside the above domains. For an arbitrary partition
of $\{1, \dots, n\}$, a similar factorization condition can be
derived from the (anti-)symmetry axiom. If
there are any fermionic fields in the theory, then
there are $\pm$-signs.

We also note that we may iterate the above factorization equation on suitable domains.
For example, if the $j$-th subset $I_j$ is itself partitioned into subsets, then
on a suitable subdomain associated with the partition, the coefficient $\C(X_{I_j})$
itself will factorize. Subsequent partitions may naturally be identified with
trees on $n$ elements $\{1, \dots, n\}$, i.e., the specification of a tree naturally
corresponds to the specification of a nested set of subsets of $\{1, \dots, n\}$.
In~\cite{HW08} and also below, a version of the above factorization property is given in terms of such
trees. However, we note that the condition given in reference~\cite{HW08} is not stated in
terms of convergent power series expansions, but instead in terms of asymptotic
scaling relations. The former seems to be more natural in the Euclidean domain.

\medskip
\noindent
\paragraph{\bf Scaling:} Let $|v_{a_1}\rangle, \dots, |v_{a_n} \rangle \in V$ be vectors with dimension
$\Delta_1, \dots, \Delta_n$ [see the decomposition  of $V$ in
eq.~\eqref{decomp}] respectively, and let $\langle v^b | \in V^*$ be an element
in the dual space of $V$ with dimension $\Delta_{n+1}$. Then the scaling degree\footnote{
The scaling degree is defined here as the infimum over all $p \in \mr$ such that
$\lim \epsilon^p C_{a_1 \dots a_n}^b(\epsilon x_1, \dots, \epsilon x_n) = 0$ for all
$(x_1, \dots, x_n) \in M_n$.}
of the $\mc$-valued distribution~\eqref{Ccompdef}
should be estimated by
\ben
sd \, C_{a_1 \dots a_n}^b \le \Delta_1 + \dots + \Delta_n - \Delta_{n+1} \, .
\een
If $v^b$ is an element of the 1-dimensional subspace of dimension-0 fields
spanned by the identity operator $\myid \in V$, if $n=2$ and if $|v_{a_1}^{} \rangle =
|v_{a_2}^\star \rangle \neq 0$, then it is required that the inequality is saturated.

\medskip
\noindent
\paragraph{\bf (Anti-)symmetry:} Let $\tau_{i-1, i} = (i-1 \,\, i)$ be the permutation
exchanging the $(i-1)$-th and the $i$-th object, which we define
to act on $V \otimes \dots \otimes V$ by exchanging the corresponding
tensor factors. Then we have
\bena
&&\C(x_1, \dots, x_{i-1}, x_i, \dots, x_n) \, \tau_{i-1,i} =
\C(x_1, \dots, x_i, x_{i-1}, \dots, x_n) \, (-1)^{F_{i-1}F_i} \\
&& F_i := \frac{1}{2} \,
id^{i-1} \otimes (id-\gamma)
\otimes id^{n-i} \, .
\eena
for all $1<i<n$. Here, the last factor is designed so that Bosonic fields
have symmetric OPE coefficients, and Fermi fields have anti-symmetric
OPE-coefficients. The last point $x_n$, and the $n$-th tensor factor
in $V\otimes \dots \otimes V$
do not behave in the same way under permutations. This is because we have
chosen to expand an operator product around the $n$-th (i.e., last) point,
and hence this point and tensor factor is
not on the same footing as the other points and tensor factors in the OPE.
The corresponding (anti-)symmetry property for
permutations involving $x_n$ is as follows. We let $t(x_1, x_n)$ be the
Taylor expansion map explained in the identity element axiom. Then we postulate
\ben
\C(x_1, \dots, x_{n-1}, x_n) \, \tau_{n-1,n} =
t(x_{n-1}, x_n) \, \C(x_1, \dots, x_n, x_{n-1}) \, (-1)^{F_{n-1}F_{n}}
\een
The additional factor of the Taylor expansion operator
$t(x_{n-1}, x_n)$ compensates for the change in the reference point.
This formula can be motivated heuristically in a similar way as
the similar formulae in the identity axiom.

\medskip
\noindent
The factorization property~\eqref{factorization} is the core property of the
OPE coefficients that holds everything together. It is clear that it imposes very stringent
constraints on the possible consistent hierarchies $(\C(x_1, x_2), \C(x_1, x_2, x_3),
\dots )$. The Euclidean invariance axiom
implies that the OPE coefficients are translation invariant, and it links the
decomposition~\eqref{decomp} of the field space into sectors of different spin to the transformation
properties of the OPE coefficients under the rotation group. The scaling property likewise
links the decomposition into sectors with different dimension to the scaling properties
of the OPE coefficients. The (anti-)symmetry property is a replacement for local
(anti-)commutativity (Einstein causality) in the Euclidean setting. Note that
we do not impose here as a condition that the familiar relation between spin and
statistics~\cite{Wightman} should hold. As we have shown in~\cite{HW08}, this may be derived
as a consequence of the above axioms in the variant considered there. Similarly,
we do not postulate any particular transformation properties under discrete
symmetries such as $C,P,T$, but we mention that one can derive the $PCT$-theorem
in this type of framework, as shown in~\cite{HPCT}. The same
result may also be proved in the present setting by very similar techniques,
but we shall not dwell upon this here.

In summary, in the following, a quantum field theory is defined as a pair
consisting of an infinite dimensional vector space $V$ with the above stated
properties, together with a hierarchy of OPE coefficients
$\C:= (\C(x_1, x_2), \C(x_1, x_2, x_3), \dots)$
with the above stated properties. It is natural to identify quantum field theories if
they only differ by a redefinition of its fields. Informally, a field redefinition
means that one changes ones definition of the quantum fields of the theory
from $\phi_a(x)$ to $\widehat \phi_a(x) = \sum_b z_a^b \phi_b(x)$, where $z_a^b$ is some matrix on field space.
The OPE coefficients of the redefined fields differ from the original ones accordingly by
factors of this matrix. We formalize this in the following definition:

\begin{defn}\label{fieldred}
Let $(V, \C)$ and $(\widehat V, \widehat \C)$ be two quantum field theories. If there exists
an invertible linear map $z: V \to \widehat V$ with the properties
\ben
z \, R(g) = \hat R(g) \, z \,, \quad z \, \gamma = \hat \gamma \, z \, , \quad
z \, \star = \hat \star \, z \, ,
\een
together with
\ben
\C(x_1, \dots, x_n) =  z^{-1} \, \widehat \C(x_1, \dots, x_n) \, z^n
\een
for all $n$, where $z^n = z \otimes \dots \otimes z$, then the two quantum field
theories are said to be equivalent, and $z$ is said to be a field redefinition.
\end{defn}

We would finally like to impose a condition that the quantum field theory $(V, \C)$ described
by the field space $V$ and the OPE coefficients $\C$ has a vacuum state.
Since we are working in a Euclidean setting here, the appropriate notion
of quantum state is a collection of Schwinger- or correlation functions, denoted as usual by
$\langle \phi_{a_1}(x_1) \cdots \phi_{a_n}(x_n) \rangle_\Omega$, where $n$ and $a_1, \dots, a_n$
can be arbitrary. These functions should be analytic functions on $M_n$ satisfying the
Osterwalder-Schrader (OS) axioms for the vacuum state $\Omega$~\cite{OS1,OS2}. They should also satisfy the
OPE in the sense that
\ben
\big\langle \phi_{a_1} (x_1) \cdots \phi_{a_n}(x_n) \big\rangle_\Omega \sim
\sum_{b} C_{a_1 \dots a_n}^b(x_1, \dots, x_n) \, \big\langle \phi_b(x_n) \big\rangle_\Omega \, .
\een
Here, the symbol $\sim$ means that the difference between the left and right side is
a distribution on $M_n$ whose scaling degree is smaller than any given number $\delta$
provided the above sum goes over all of the finitely many
fields $\phi_b$ whose dimension is smaller than some number $\Delta = \Delta(\delta)$.
The OS-reconstruction theorem then guarantees that the theory can
be continued back to Minkowski spacetime, and that the fields can be represented as
linear operators on a Hilbert space $\H$ of states. One may want to
impose only the weaker condition that there exist {\em some} quantum state for the
quantum field theory described by $(\C, V)$. In that case, one would postulate the
existence of a set of Schwinger functions satisfying all of the
OS-axioms except those involving statements about the invariance under the Euclidean group.
Such a situation is of interest in theories with unbounded potentials where a
vacuum state is not expected to exist, but where the OPE might nevertheless exist.

It is clear that the existence of a vacuum state (or in fact, just any quantum state)
satisfying the OS-axioms is a potentially new restriction on the OPE coefficients.
We will not analyze here the nature of these restrictions, as our focus is on the algebraic
constraints satisfied by the OPE-coefficients. We only note here that the condition of
OS-positivity is not satisfied in some systems in statistical mechanics, and it is also
not satisfied in gauge theories before the quotient by the BRST-differential is taken~(see
sec.~\ref{hochschild}). These systems on the other hand do satisfy an OPE in a suitable sense.
Thus, one would expect that the existence of
a set of correlation functions satisfying the full set of
OS-axioms is a genuinely new restriction\footnote{
Consequences of OS-positivity have been analyzed in the context of partial wave expansions~\cite{Rehren1,Rehren2},
and also in the framework of~\cite{Mack1}.
} on
the allowed theory, which one might want to drop in some cases.

\section{Coherence theorem}\label{coherence}

In the last section we have laid out our definition of a quantum field theory in
terms of a collection of operator product coefficients. The key condition that
these should satisfy is the factorization property~\eqref{factorization}. It is
clear that these conditions should impose a set of very stringent constraints upon the
coefficients $\C(x_1, \dots, x_n)$ for $n \ge 2$. In this section, we will analyze
these conditions and show that, in a sense, all of these constraints may be thought
of encoded in the first non-trivial one arising at $n=3$ points. We shall
refer to this type of result as a "coherence theorem", because it means that
all the factorization constraints are coherently described by a single condition
in the precise sense explained below.

Before we describe our result in detail, we would
like to put it into perspective by drawing a parallel to an analogous
result valid for ordinary algebras. Let $\A$ be a finite-dimensional algebra.
The key axiom for an algebra is the associativity condition, stating that
\ben\label{aassoc}
(AB)C = A(BC) \quad \text{for all $A,B,C \in \A$.}
\een
Written somewhat differently, if we write the product as
$m(A,B) = AB$ with $m$ a linear map $m: \A \otimes \A \to \A$, then
in a tensor product notation similar to the one used above
in context of the OPE, the associativity condition is equivalent to
\ben\label{massoc}
m(id \otimes m) = m(m \otimes id) \, ,
\een
where the two sides of the above equation are now maps $\A \otimes \A \otimes \A \to \A$.
An elementary result for algebras is that there do not arise any
further constraints on the product $m$ from "higher associativity conditions" such
as for example
\ben\label{ahigherass}
(AB)(CD) = (A(BC))D \quad \text{for all $A,B,C,D \in \A$.}
\een
Indeed, it is not difficult to prove this identity by successively
applying eq.~\eqref{aassoc}, and this can be generalized
to prove all possible higher associativity identities.
The associativity condition~\eqref{aassoc} is analogous to
the consistency conditions for the OPE coefficients arising from the
the factorization constraint~\eqref{factorization}
for three points. Moreover, the higher order associativity conditions~\eqref{ahigherass}
are analogous
to the conditions that arise from the factorization constraint for more than three
points. Thus, our coherence theorem is analogous to the above statement for
ordinary algebras that there are no higher order associativity constraints which
are not already automatically satisfied on account of the
standard associativity condition~\eqref{aassoc}.

Let us now describe our coherence result in more detail.
For $n=3$ points, there are three partitions of the set $\{1, 2, 3\}$ leading
to three corresponding non-trivial factorization conditions~\eqref{factorization}, namely\footnote{
Note that, in our formulation of the factorization condition, there is an ordering
condition on the partitions. Here we mean more precisely all conditions that can be obtained
by combining this with the symmetry axiom, which will give conditions for arbitrary orderings.}
$\T_3:=\{ \{1, 2\}, \{3\} \}$, $\T_2 := \{ \{1, 3\}, \{2\} \}$, and $\T_1:=
\{ \{2, 3\}, \{1\} \}$. The corresponding domains on which the factorization
identities are valid are given respectively by
\bena\label{threet}
\D[\T_1] &=& \{(x_1, x_2, x_3) \mid r_{23} < r_{13} \} \, ,\\
\D[\T_2] &=& \{(x_1, x_2, x_3) \mid r_{13} < r_{23} \} \, ,\\
\D[\T_3] &=& \{(x_1, x_2, x_3) \mid r_{12} < r_{23} \} \, .
\eena
Clearly, the first two domains have no common points, but they both have an open,
non-empty intersection with the third domain. Thus, on each of
these intersections, we have two factorizations of the OPE coefficient
$\C(x_1, x_2, x_3)$ according to eq.~\eqref{factorization}.
These must hence be equal. Thus,
we conclude that
\ben\label{Cassoc}
\C(x_2, x_3) \Big( \C(x_1, x_2) \otimes id \Big) = \C(x_1, x_3) \Big( id \otimes \C(x_2, x_3) \Big) \,
\een
on the intersection $\D[\T_1] \cap \D[\T_3]$ [that is, the set
$\{r_{12}<r_{23}<r_{13}\}$] and a similar relation must hold
on the intersection $\D[\T_2] \cap \D[\T_3]$. However, the latter relation is
can also be derived from eq.~\eqref{Cassoc} by the
symmetry axiom for the OPE coefficients stated in the previous section,
\ben\label{add1}
\C(x_1, x_2) = t(x_1, x_2) \C(x_2, x_1) \tau_{1,2} \,
\een
and the relation
\ben\label{add2}
\C(x_1, x_3) = \C(x_2, x_3)\Big( t(x_1, x_2) \otimes id \Big)
\een
for $r_{12} < r_{23}$.
Thus, for three points, essentially the only independent
consistency condition is eq.~\eqref{Cassoc}. In component form, this condition
was given above in eq.~\eqref{assoccomp}.

The consistency condition~\eqref{Cassoc} is analogous to the
associativity condition~\eqref{massoc} for the product in
an ordinary algebra. By analogy to an ordinary algebra, we may
hence ask whether there are any further constraints on $\C(x_1, x_2)$ arising
from the higher order factorization equations~\eqref{factorization} with
$n \ge 4$. As we will now show, this is not the case. We also show that, as in
an ordinary algebra, the coefficients $\C(x_1, \dots, x_n)$ analogous to a product
of $n$ factors are completely determined by the coefficient $\C(x_1, x_2)$ analogous
to a product with two factors.

Our first task is to write down all factorization conditions involving only the
coefficients $\C(x_1, x_2)$. For this, it is useful to employ the language of rooted
trees. One way to describe a rooted tree on $n$ elements $\{1, \dots, n\}$ is
by a set $\{S_1, \dots, S_k\}$ of nested subsets $S_i \subset \{1, \dots, n\}$.
This is a family of subsets with the property that each set $S_i$ is either contained
in another set of the family, or disjoint from it. The set $\{1, \dots, n\}$ is
by definition not in the tree, and is referred to as the root.
The sets $S_i$ are to be thought of as the
nodes of the tree, and a node is connected by branches to all those nodes
that are subsets of $S_i$ but not proper subsets of any element of the tree other
than $S_i$. The leaves are those nodes that
themselves do not possess any other set $S_i$ in the tree and are
given by the singleton sets $S_i = \{i\}$. If $\T$ is a tree
on $n$ elements of a set, then we also denote by $|\T|$ the elements
of this set. Let $\T$ be a tree upon $n$
elements of the form $\T = \{ \T_1, \dots, \T_r \}$, where each $\T_i$ is
itself a tree on a proper subset of $\{1, \dots, n\}$, so that $|\T_1| \cup \dots \cup
|\T_r| = \{1, \dots, n\}$ is a partition into disjoint subsets.
We define an open, non-empty domain of $M_n$ for such trees recursively
by
\bena\label{dtdef}
\D[\T] &=&
\bigg\{
(x_1, \dots, x_n) \in M_n \mid X_{|\T_1|} \in \D[\T_1], \dots,
X_{|\T_r|} \in \D[\T_r]; \non\\
&&{\rm min} \, d(X_{\{m_1, \dots, m_r\}}) > {\rm max} \, (d(X_{|\T_1|}), \dots, d(X_{|\T_r|}))
\bigg\}
\, ,
\eena
where $m_i$ is the maximum element upon which the tree $\T_i$ is
built, and where we are using the same notations $d(X_I)$ and
$X_I = (x_i)_{i \in I}$ as above for any subset $I \subset \{1, \dots, n\}$.
If $\T_i$ are the trees with only a single node apart from the leaves,
then the above domain is identical with the domain defined above in the
factorization axiom~\eqref{factorization}, see eq.~\eqref{domaindef} with
$I_i$ in that definition given by the elements of the $i$-th subtree $\T_i$.
Otherwise, it is a proper open subset of that domain. In
any case, the factorization identity~\eqref{factorization} holds on $\D[\T]$.
However, we may now iterate the factorization identity, because the
factors $\C(X_{|\T_i|})$ now themselves factorize on $\D[\T]$, given
that $X_{|\T_i|} \in \D[{\T_i}]$. We apply the factorization condition
to this term again, and continuing this way, we get a nested factorization identity on
each of the above domains $\D[\T]$.

To write down these identities in a reasonably compact way,
we introduce some more notation. If $S \in \T$, we write
$\ell(1), \dots, \ell(j) \subset_\T S$ if $\ell(1), \dots, \ell(j)$ are the
branches descending from $S$ in the tree $\T$. We write
$m_i$ for the largest element in the sets $\ell(i)$, and we assume that the branches
have been ordered in such a way that $m_1 < \dots < m_j$.
As above in eq.~\eqref{Ccompdef}, we let
$C_{a_1 \dots a_n}^b(x_1, \dots, x_n)$ be the
basis components of the linear maps $\C(x_1, \dots, x_n): V^{\otimes n} \to V$.
Then, for each tree $\T$ on $\{1, \dots, n\}$, the
following factorization identity holds on the domain $\D[\T]$:
\ben\label{treefactor}
C_{a_1 \dots a_n}^b(x_1, \dots, x_n) =
\sum_{a_S: S \in \T} \left( \prod_{S: \ell(1), \dots, \ell(j) \subset_\T S}
C^{a_S}_{a_{\ell(1)} \dots a_{\ell(j)}}
(x_{m_1}, \dots, x_{m_j}) \right) \, .
\een
Here, the sums are over all $a_S$ with $S$ a subset in the tree not equal to $\{1\}, \dots, \{n\}$
respectively
$\{1, \dots, n\}$ . For these sets, we define
$a_{\{1\}} := a_1, \dots, a_{\{n\}} := a_n$ respectively
$a_{\{1, \dots, n\}} := b$. The nested infinite sums are carried
out in the hierarchical order determined by the tree, with the sums corresponding
to the nodes closest to the leaves first. If $\T$ is a binary
tree, i.e., one where precisely two branches descend from each node, then
the above factorization formula expresses the $n$-point
OPE coefficient $\C(x_1, \dots, x_n)$ in
terms of products of the 2-point coefficient in the open domain $\D[\T] \subset M_n$.
Since $\C(x_1, \dots, x_n)$ is by assumption analytic in the open, connected domain $M_n$,
and since an analytic function on a connected domain
is uniquely determined by its restriction to an open set, we have the
following simple proposition:

\begin{prop}\label{proposition1}
The $n$-point OPE-coefficients $\C(x_1, \dots, x_n)$ are uniquely determined
by the 2-point coefficients $\C(x_1, x_2)$. In particular, if two quantum
field theories have equivalent 2-point OPE coefficients [see the previous section],
then they are equivalent.
\end{prop}

We next ask whether the factorization condition~\eqref{treefactor}
for binary trees $\T$ imposes any further restrictions on $\C(x_1, x_2)$ apart
from~\eqref{Cassoc}. For this, consider for any binary tree $\T$ the expression
\ben\label{ftdef}
(f_\T)^b_{a_1 \dots a_n}(x_1, \dots, x_n) :=
\sum_{a_S: S \in \T} \left( \prod_{S: \ell(1), \ell(2) \subset_\T S}
C^{a_S}_{a_{\ell(1)} a_{\ell(2)}}
(x_{m_1}, x_{m_2}) \right)
\een
defined on the domain $\D[\T]$.
Thus, $f_\T(x_1, \dots, x_n)$ is the expression for
$\C(x_1, \dots, x_n)$ in the factorization condition~\eqref{treefactor}
for the binary tree $\T$. This factorization condition hence implies that
$f_\T$ can be analytically continued to an analytic function on $M_n$
(denoted again by $f_\T$), and that this $f_\T$ is in fact independent
of the choice of the binary tree $\T$. In order to see to what kinds of
constraints this puts on the 2-point OPE coefficients $\C(x_1, x_2)$, let us now
pretend we only knew that the sums converge in
eq.~\eqref{ftdef}, that they define
an analytic function $f_\T$ on $\D[\T]$, and that this can be analytically
continued to $M_n$, for all $n$ and all binary trees on $n$ elements.
In particular, for the sake of the argument,
let us {\em not} assume that the $f_\T$ coincide for different binary trees $\T$,
except in the case $n=3$. In this case, the assumption that $f_\T$ coincide for the
three binary trees and corresponding domains~\eqref{threet} is equivalent
to the assumption of associativity for three points [see eq.~\eqref{Cassoc}]
and the symmetry and normalization conditions~\eqref{add1},\eqref{add2},
and we want to assume this condition.

\begin{figure}
\begin{center}
\includegraphics[width=6.5in]{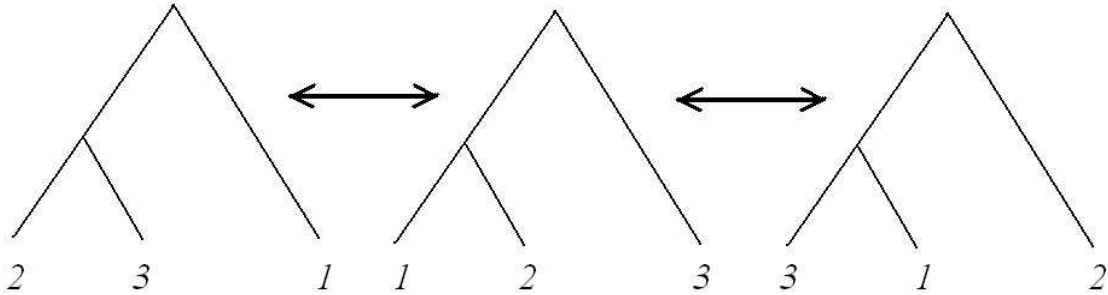}
\end{center}
\caption{A graphical representation of the associativity condition. The double arrows indicate that
the domains $\D[\T_i]$ represented by the respective trees have a common intersection, and that on this
intersection, the OPE's represented by the respective trees coincide. Note that the double arrows
are not a transitive relation: The domains associated with left- and rightmost tree have empty
intersection.}
\label{fig1}
\end{figure}

\begin{figure}
\begin{center}
\includegraphics[width=3.5in]{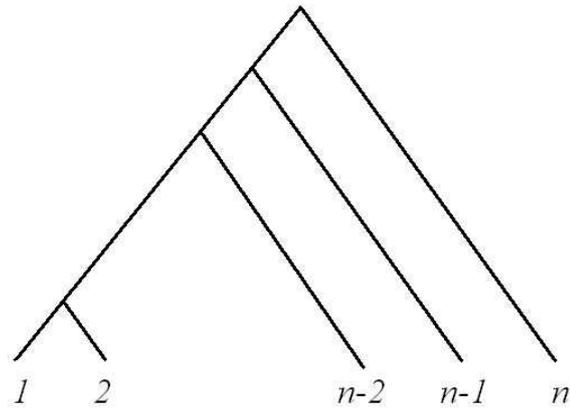}
\end{center}
\caption{The reference tree $\bf S$.}
\label{fig2}
\end{figure}

We will now show that these assumptions in fact imply that all $f_\T$ coincide
for all binary trees $\T$. In this sense, there are no further consistency conditions
on $\C(x_1, x_2)$ beyond those for three points. The proof of this statement is
not difficult, and is in fact very similar to the proof of the corresponding statement
for ordinary algebras. The argument is most easily presented
graphically in terms of trees. For $n=3$, we graphically
present the assumption that all $f_\T$ agree
for the three trees associated with three elements as fig.~\ref{fig1}.
In this figure, each tree symbolizes the corresponding expression $f_\T$, and an arrow
between two trees means the following relation: (i) the intersection of the
corresponding domains [see eq.~\eqref{threet}] is
not empty, and (ii) the expressions coincide on that intersection. Because
the $f_\T$ are analytic, any such relation implies that the corresponding $f_\T$'s
in fact have to coincide everywhere on $M_n$.
Now consider $n>3$ points, and let $\T$ be an arbitrary
tree on $n$ elements. The goal is to present a sequence of trees $\T_0, \T_1, \dots, \T_r$
of trees such that $\T_0 = \T$, and such that $\T_r = \S$ is the "reference tree"
\ben
\S = \{ \{n\}, \{ n-1, n\}, \{n-2, n-1, n\}, \dots, \{1, 2, \dots n\} \}
\een
which is drawn in fig.~\ref{fig2}. The sequence should have the further property that
for each $i$, there is a relation as above between $\T_i$ and $\T_{i-1}$. As we
have explained, this would imply that $f_\T = f_\S$, and hence that all
$f_\T$'s are equal.

\begin{figure}
\begin{center}
\includegraphics[width=7.0in]{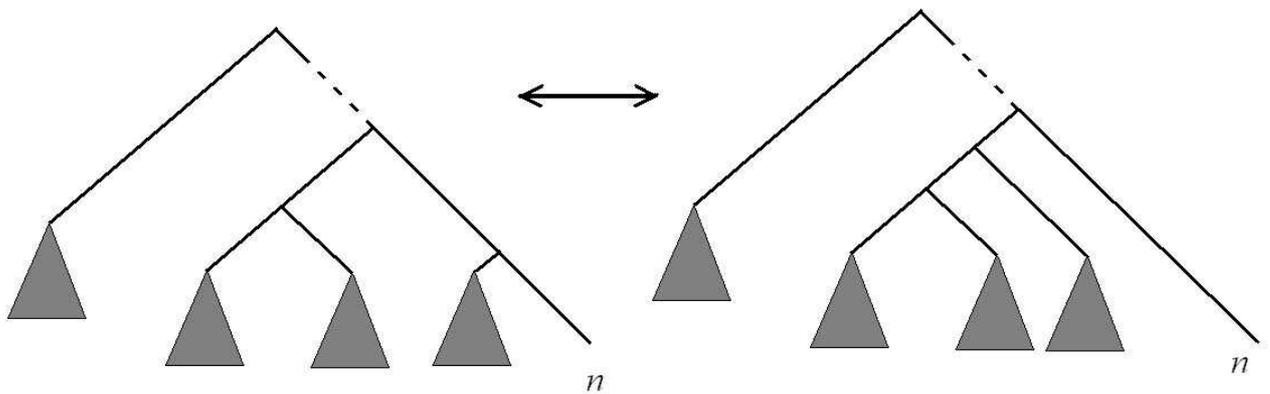}
\end{center}
\caption{An elementary manipulation. The shaded triangles represent subtrees whose form is not relevant.}
\label{fig3}
\end{figure}

\begin{figure}
\begin{center}
\includegraphics[width=7.5in]{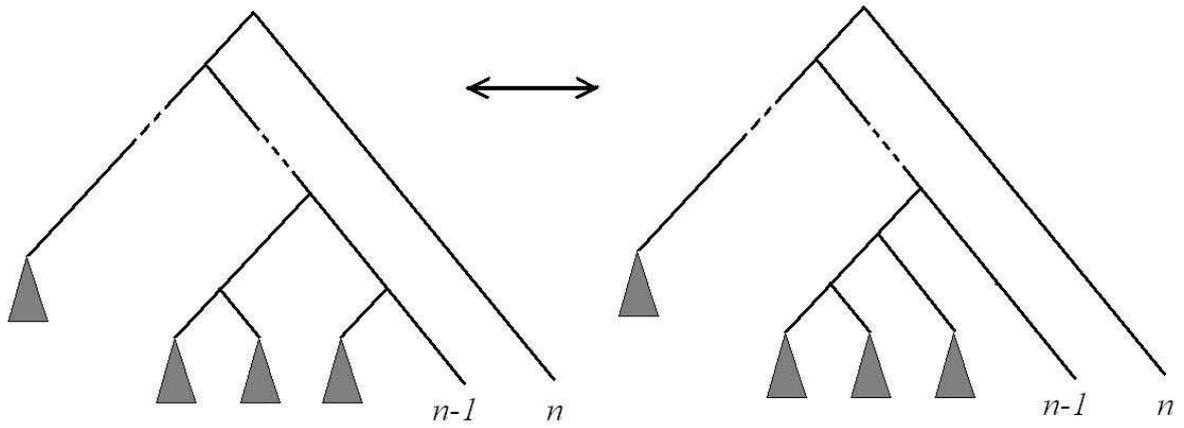}
\end{center}
\caption{Another elementary manipulation.}
\label{fig4}
\end{figure}

\begin{figure}
\begin{center}
\includegraphics[width=4.5in]{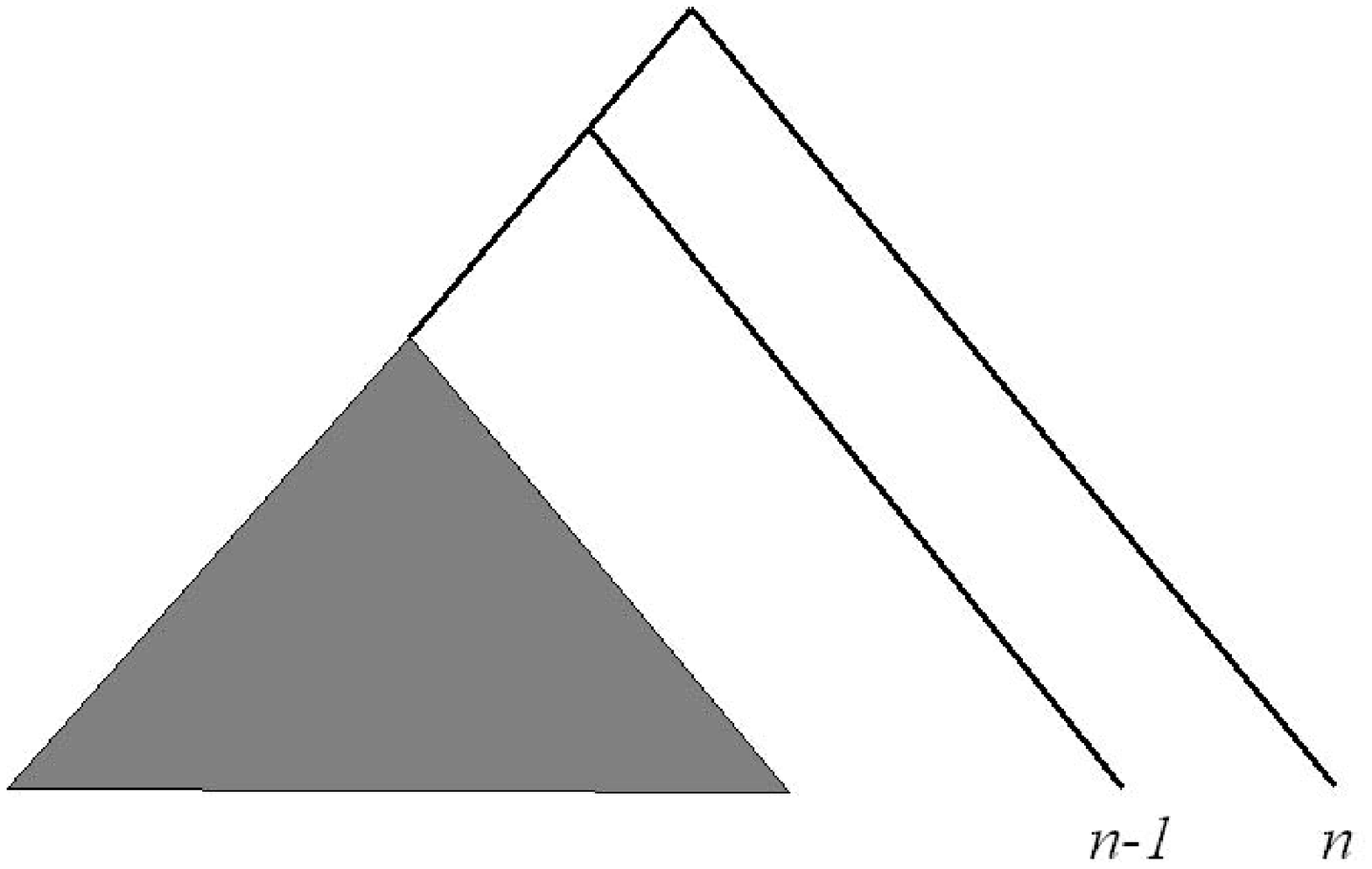}
\end{center}
\caption{The tree $\T_{r_2}$.}
\label{fig5}
\end{figure}

We now construct the desired sequence of trees inductively. We first write the binary
tree $\T=\T_0$ as the left tree in fig.~\ref{fig3}, where the shaded regions again represent subtrees whose
particular form is not relevant. The next tree $\T_1$ is given by the right tree in
fig.~\ref{fig3}. We claim that there is a relation as above between these trees. In fact, it is easy to
convince oneself that the corresponding domains $\D[\T_0]$ and $\D[\T_1]$ have a non-empty
intersection. Secondly, because these trees differ by an elementary manipulation
as in fig.~\ref{fig1}, it is not difficult to see that the three-point consistency condition
implies that the corresponding expressions $f_{\T_0}$ and $f_{\T_1}$ coincide on
(at least an open subset of) $\D[\T_0] \cap \D[\T_1]$. Being analytic, they must hence coincide everywhere.
We now repeat this kind of process until
we arrive at the left tree $\T_{r_1}$ in fig.~\ref{fig4}. This tree has
the property that the $n$-th leaf is directly connected to the root. We change this
tree to the right tree in fig.~\ref{fig4}, again verifying that there is indeed the desired relation
between these trees. We repeat this step again
until we reach the tree $\T_{r_2}$ given in fig.~\ref{fig5}. It is clear now that this can be continued
until we have reached the tree $\S$ in fig.~\ref{fig2}.

We summarize our finding in the following theorem:

\begin{thm}
("Coherence Theorem")
For each binary tree $\T$, let $f_\T$ be defined by eq.~\eqref{ftdef} on the domain $\D[\T]$
as a convergent power series expansion, and assume that $f_\T$ has
an analytic extension to all of $M_n$. Furthermore, assume that the associativity
condition~\eqref{Cassoc} and symmetry and normalization conditions~\eqref{add1}, \eqref{add2}
hold, i.e. that all $f_\T$ coincide for trees
with three leaves. Then $f_\T = f_\S$ for any pair of binary trees $\S, \T$.
\end{thm}

\section{Perturbations and Hochschild cohomology}\label{perturbations}

Suppose we are given a quantum field theory in terms of
OPE-coefficients as described in sec.~\ref{axiomatic}. In this section we discuss the question
how to describe perturbations of such a quantum field theory.
According to our definition of a quantum field theory, a
perturbed quantum field theory should correspond to a
perturbation series in some parameter $\lambda$ for the OPE coefficients.
Because our axioms for the OPE coefficients imply constraints--especially
the factorization axiom--the perturbations of the coefficients
will also have to satisfy corresponding constraints. In this
section, we will show that these constraints are of a
cohomological nature.

As we have discussed, our definition of quantum field theory
is algebraic. In fact, as argued in sec.~\ref{coherence}, up to
technicalities related to the convergence of various series,
the constraints on the OPE coefficients
can be formulated in the form of an "associativity condition" for the
2-point OPE coefficients only, see eq.~\eqref{Cassoc}. Consequently,
the perturbed 2-point OPE coefficients will also have to satisfy
a corresponding perturbed version of this constraint, and this
is in fact essentially the only constraint. It is this
perturbed version of the associativity condition that we will
discuss in this section.

Our discussion is in close parallel to the well-known
characterization of perturbations ("deformations") of an
ordinary finite dimensional algebra, an analogy which we have already
emphasized in another context above. We therefore begin by recalling
the basic theory of deformations
of finite-dimensional algebras~\cite{Gerstenhaber, Happel}.
Let $\A$ be a finite-dimensional algebra (over $\mc$, say), whose product
we denote as usual by $\A \otimes \A \to \A, A \otimes B \mapsto AB$.
A deformation of the algebra is a 1-parameter family of
products $A \otimes B \mapsto A \bullet_\lambda B$, where
$\lambda \in \mr$ is a smooth deformation parameter. The product
$A \bullet_0 B$ should be the original product $AB$, but
for non-zero $\lambda$, we have a new product on $\A$---or alternatively
on the ring of formal power series $\mc((\lambda)) \otimes \A$ if we merely
consider perturbations in the sense of formal power series. This new product
must satisfy the associativity law, which imposes a strong constraint.
If we denote the $i$-th order perturbation of the product by
\ben
m_i(A, B) = \frac{1}{i!} \, \frac{d^i}{d\lambda^i} A \bullet_\lambda B \Bigg|_{\lambda=0} \, ,
\een
then the associativity condition implies to first order that
we should have
\ben
m_0(\id \otimes m_1) - m_0(m_1 \otimes \id) + m_1(\id \otimes m_0) -
m_1(m_0 \otimes \id) = 0 \, ,
\een
as a map $\A \otimes \A \otimes \A \to \A$, in an obvious tensor
product notation. $m_0(A,B) = AB$ is the original product on $\A$. Similar conditions
arise for the higher derivatives $m_i$ of the new product. These may be
written for $i \ge 2$ as
\bena
&&m_0(\id \otimes m_i) - m_0(m_i \otimes \id) + m_i(\id \otimes m_0) -
m_i(m_0 \otimes \id) \non\\
&=& -\sum_{j=1}^{i-1} m_{i-j}(\id \otimes m_j) - m_{i-j}(m_j \otimes \id)
\, .
\eena
Actually, we want to exclude the trivial case that the new product was
obtained from the old one by merely a $\lambda$-dependent redefinition of the
generators of $\A$. Such a redefinition may be viewed as a 1-parameter family of
invertible linear maps $\alpha_\lambda:
\A \to \A$, and the corresponding trivially deformed product is
\ben\label{trivial}
A \bullet_\lambda B =
\alpha_\lambda^{-1}\Big[\alpha_\lambda^{}(A) \alpha_\lambda^{}(B) \Big] \, .
\een
In other words, $\alpha_\lambda$ defines an isomorphism between
$(\A, \bullet_0)$ and $(\A, \bullet_\lambda)$, meaning that the latter
should not be regarded as a new algebra.
The trivially deformed product is given to first order by
\ben
m_1 = m_0(\id \otimes \alpha_1) + m_0(\alpha_1 \otimes \id) - \alpha_1 m_0 \, ,
\een
with similar formulas for $m_i$, where $\alpha_i = \frac{1}{i!}\, \frac{d^i}{d\lambda^i} \alpha_\lambda |_{\lambda=0}$.

The above conditions for the $i$-th order deformations of
an associative product have a useful and elegant cohomological interpretation~\cite{Gerstenhaber}.
To give this interpretation, consider the linear space $\Omega^n(\A)$ of
all linear maps $\psi_n: \A \otimes \dots \otimes  \A \to \A$, and define a linear operator
$d: \Omega^n \to \Omega^{n+1}$ by the formula
\bena
(d \psi_n)(A_1, \dots, A_{n+1}) &=& A_1 \psi_n(A_2, \dots, A_{n+1}) -
(-1)^n \psi_n(A_1, \dots, A_n)A_{n+1} \non\\
&& +\sum_{j=1}^n (-1)^j \psi_n(A_1, \dots, A_{j} A_{j+1}, \dots, A_{n+1}) \, .
\eena
It may be checked using the associativity law for the original product on the algebra $\A$ that
$d^2 = 0$, so $d$ is a differential with a corresponding cohomology complex. This complex
is called the Hochschild complex, see e.g.~\cite{Connes}. More precisely, if $Z^n(\A)$ is the space of
all closed $\psi_n$, i.e., those satisfying $d \psi_n=0$, and $B^n(\A)$ the space of
all exact $\psi_n$, i.e., those for which $\psi_n = d \psi_{n-1}$ for some
$\psi_{n-1}$, then the $n$-th Hochschild cohomology
$HH^n(\A)$ is defined as the quotient $Z^n(\A)/B^n(\A)$.
The first order associativity condition may now be viewed as saying that $d m_1 = 0$, or
$m_1 \in Z^2(\A)$. Furthermore, if the new product just arises from a trivial
redefinition of the generators in the sense of~\eqref{trivial}, then
it follows that $m_1 = d \alpha_1$, so $m_1 \in B^2(\A)$ in that case.
Thus, the non-trivial first order perturbations $m_1$ of the algebra
product can be identified with the non-trivial classes $[m_1] \in HH^2(\A)$.
In particular, non-trivial deformations may only exist if $HH^2(\A) \neq 0$.
Let us assume a non-trivial first order perturbation exists, and let us try
to find a second order perturbation. We view the right side of the second
order associativity condition as an element $w_2 \in \Omega^3(\A)$,
and we compute that
$d w_2 = 0$, so $w_2 \in Z^3(\A)$. Actually, the left side of the second order
associativity condition is just $d m_2 \in B^3(\A)$ in our cohomological notation,
so if the second order associativity condition is to hold, then $w_2$ must in fact
be an element of $B^3(\A)$, or equivalently, the class $[w_2] \in HH^3(\A)$ must
vanish. If it does not define the trivial class---as may only happen if
$HH^3(\A) \neq 0$ itself is non-trivial---then there is an obstruction to
lift the perturbation to second order.
If there is no obstruction at second order, we continue to third order,
with a corresponding potential obstruction $[w_3] \in HH^3(\A)$, and so on.
In summary, the space of non-trivial perturbations corresponds to
elements of $HH^2(\A)$, while the obstructions lie in $HH^3(\A)$.

We now show how to give a similar characterization of perturbations of a
quantum field theory.
According to our definition of a quantum field theory
given in sec.~\ref{axiomatic}, a quantum field theory is defined by the set of
its OPE-coefficients with certain properties. Furthermore, as
argued in sec.~\ref{coherence}, all higher $n$-point operator product coefficients
are uniquely determined by the 2-point coefficients $\C(x_1, x_2)$.
Furthermore, we argued that, up to technical assumptions about
the convergence of the series~\eqref{ftdef}, the key constraints
on the OPE coefficients for $n$ points are encoded in the
associativity constraint~\eqref{Cassoc} for the 2-point coefficient,
which we repeat for convenience:
\ben\label{maincondition}
\C(x_2, x_3)\Big(\C(x_1, x_2) \otimes \id \Big) - \C(x_1, x_3)\Big(
\id \otimes \C(x_2, x_3) \Big) = 0 \quad
\text{for $r_{12} < r_{23} < r_{13}$.}
\een
We ask the question when it is possible to find a 1-parameter deformation $\C(x_1, x_2; \lambda)$ of
these coefficients by a parameter $\lambda$ so that the
associativity condition continues to hold, at least in the
sense of formal power series in $\lambda$. Actually, the analogues of
the symmetry condition~\eqref{add1}, the normalization condition~\eqref{add2}, the
hermitian conjugation, the Euclidean invariance, and the unit axiom should
hold as well for the perturbation. However, these conditions are much more
trivial in nature than~\eqref{maincondition}, because these conditions
are linear in $\C(x_1, x_2)$. These conditions could therefore easily be included in
our discussion, but would distract from the main point. For the rest of this section,
we will therefore discuss the implications of the associativity condition~\eqref{maincondition}
for the perturbed OPE-coefficients.

As we shall see now,
such perturbations can again be characterized in a cohomological framework
similar to the one given above. As above, we will presently define a
linear operator $b$ which defines the cohomology in question. The definition
of this operator will implicitly involve infinite sums [as our
associativity condition~\eqref{maincondition}], and such sums are
typically only convergent on certain domains. It is therefore necessary to get a set of domains
that will be stable under the action of $b$ and that is suitable
for our application. Many such domains can be defined, and
correspondingly different rings are obtained. For simplicity and definiteness,
we consider the non-empty, open domains of $(\mr^D)^n$ defined by
\ben\label{Fndef}
\F_n = \{(x_1, \dots, x_n) \in M_n; \,\,\, r_{1 \, i-1} < r_{i-1 \, i} < r_{i-2 \, i}
< \dots < r_{1i}, \,\,\, 1<i\le n \} \subset M_n \, .
\een
These domains also have a description in terms of the domains $\D[\T]$ defined
above in eq.~\eqref{dtdef}, but we will not need this here. Note that the associativity
condition~\eqref{maincondition} holds on the domain $\F_3 = \{r_{12} < r_{23} < r_{13}\}$.

We define $\Omega^n(V)$ to be the set of all holomorphic functions
$f_n$ on the domain $\F_n$ that are valued in the linear maps~\footnote{The
same remark as in footnote 1 applies here.}
\ben
f_n(x_1, \dots, x_n): V \otimes \dots \otimes V \to V, \quad (x_1, \dots, x_n) \in \F_n \, .
\een
We next introduce a boundary operator $b: \Omega^n(V) \to \Omega^{n+1}(V)$ by the formula
\bena\label{bfndef}
&&(b f_n)(x_1, \dots, x_{n+1}) :=
\C(x_1, x_{n+1})(\id \otimes f_n(x_2, \dots, x_{n+1})) \non\\
&&+ \sum_{i=1}^n (-1)^i f_n(x_1, \dots, \widehat x_i, \dots, x_{n+1})(
\id^{i-1} \otimes \C(x_i, x_{i+1}) \otimes \id^{n-i}) \non\\
&&+(-1)^{n+1} \, \C(x_n, x_{n+1})(f_n(x_1, \dots, x_n) \otimes \id) \, .
\eena
Here $\C(x_1, x_2)$ is the OPE-coefficient of the undeformed theory
and a caret means omission.
The definition of $b$ involves a composition of $\C$ with $f_n$, and
hence, when expressed in a basis of $V$, implicitly involves an
infinite summation over the basis elements of $V$. We must therefore
assume here (and in similar formulas in the following) that these
sums converge on the set of points $(x_1, \dots, x_{n+1})$ in the domain $\F_{n+1}$.
Thus, when we write $bf_n$, it is understood that $f_n \in \Omega^n(V)$ is
in the domain of $b$. We now have the following lemma:

\begin{lemma}
The maps $b$ is a differential, i.e., $b^2f_n = 0$ for
$f_n$ in the domain of $b$ such that $bf_n$ is also in the domain of $b$.
\end{lemma}

\medskip
\noindent
{\em Proof:} The proof is essentially a straightforward computation. Using the definition of
$b$, we have
\bena\label{bbfn}
&&b(b f_n)(x_1, \dots, x_{n+2}) =
\C(x_1, x_{n+2}) (\id \otimes b f_n(x_2, \dots, x_{n+2})) \non\\
&&+ \sum_{i=1}^{n+1} (-1)^i b f_n(x_1, \dots, \widehat x_i, \dots, x_{n+2})(
\id^{i-1} \otimes \C(x_i, x_{i+1}) \otimes \id^{n+1-i}) \non\\
&&+(-1)^{n+2} \C(x_{n+1}, x_{n+2}) (b f_n(x_1, \dots, x_{n+1}) \otimes \id) \, .
\eena
Substituting the definition of $b$ again then gives, for the first term on
the right side
\bena
&&=\C(x_1, x_{n+2})[\id \otimes \C(x_2, x_{n+2})(\id \otimes f_n(x_3, \dots, x_{n+2}))]\non\\
&&\C(x_1, x_{n+2})[\id \otimes \sum_{k=2}^{n+1} (-1)^{k-1} f_n(x_2, \dots, \widehat x_k,
\dots, x_{n+2})(\id^{k-2} \otimes \C(x_k, x_{k+1}) \otimes \id^{n-k+1})] \non\\
&&+ (-1)^{n+1} \C(x_1, x_{n+2})[\id \otimes \C(x_{n+1}, x_{n+2})(f_n(x_2, \dots, x_{n+1}) \otimes \id)]
\, .
\eena
Substituting the definition of $b$ into the third term on the right side
of eq.~\eqref{bbfn} gives
\bena
&&=(-1)^{n} \C(x_{n+1}, x_{n+2})[\C(x_1, x_{n+1})(\id \otimes f_n(x_2, \dots, x_{n+1})) \otimes id] \non\\
&&+(-1)^{n} \C(x_{n+1}, x_{n+2})[ \sum_{i=1}^n (-1)^i f_n(x_1, \dots, \widehat x_i, \dots, x_{n+1})
(\id^{i-1} \otimes \C(x_i, x_{i+1}) \otimes \id^{n-i}) \otimes id] \non\\
&&-\C(x_{n+1}, x_{n+2})[\C(x_n, x_{n+1}) (f_n(x_1, \dots, x_n) \otimes \id) \otimes id] \, .
\eena
Substituting the definition of $b$ into the second term on the right side
of eq.~\eqref{bbfn} gives the following terms
\bena
&&= \sum_{i=2}^{n+1} (-1)^i \C(x_1, x_{n+2})[\id \otimes f_n(x_2, \dots, \widehat x_i, \dots, x_{n+2})
(\id^{i-1} \otimes \C(x_i, x_{i+1}) \otimes \id^{n+1-i})] \non\\
&&
-\C(x_2, x_{n+2}) (\id \otimes f_n(x_3, \dots, x_{n+2}))(\C(x_1, x_2) \otimes \id^n)
\non\\
&& + \sum_{i=1}^{n} (-1)^{i+n+1}
\C(x_{n+1}, x_{n+2})[(f_n(x_1, \dots, \widehat x_i, \dots, x_{n+1}) \otimes \id)
(\id^{i-1} \otimes \C(x_i, x_{i+1}) \otimes \id^{n-i+1})] \non\\
&&+\C(x_n, x_{n+2}) (f_n(x_1, \dots, x_{n}) \otimes \id)(\id^n \otimes \C(x_{n+1}, x_{n+2})) \non\\
&&+\sum_{k=2}^n \sum_{i=1}^{k-1} (-1)^{k+i} f_n(x_1, \dots, \widehat x_i, \dots, \widehat x_{k+1},
\dots, x_{n+2}) \circ \non\\
&& \quad \circ (\id^{k-1} \otimes \C(x_{k+1}, x_{k+2}) \otimes \id^{n-k})
  (\id^{i-1} \otimes \C(x_{i}, x_{i+1}) \otimes \id^{n-i+1}) \non\\
&&+\sum_{k=1}^{n-1} \sum_{i=k+2}^{n+1} (-1)^{k+i} f_n(x_1, \dots, \widehat x_k, \dots, \widehat x_{i},
\dots, x_{n+2}) \circ \non\\
&& \quad \circ (\id^{k-1} \otimes \C(x_{k}, x_{k+1}) \otimes \id^{n-k})
  (\id^{i-1} \otimes \C(x_{i}, x_{i+1}) \otimes \id^{n-i+1}) \non\\
&&-\sum_{k=1}^n   f_n(x_1, \dots, \widehat x_k, \widehat x_{k+1},
\dots, x_{n+2}) \circ \non\\
&& \quad \circ (\id^{k-1} \otimes \C(x_{k}, x_{k+2}) \otimes \id^{n-k})
  (\id^{k} \otimes \C(x_{k+1}, x_{k+2}) \otimes \id^{n-k}) \non\\
  &&+\sum_{k=1}^n  f_n(x_1, \dots, \widehat x_k, \widehat x_{k+1},
\dots, x_{n+2}) \circ \non\\
&& \quad \circ (\id^{k-1} \otimes \C(x_{k+1}, x_{k+2}) \otimes \id^{n-k})
  (\id^{k-1} \otimes \C(x_{k}, x_{k+1}) \otimes \id^{n-k+1}) \, .
\eena
We now add up the expressions that we have obtained, and we use the
associativity condition eq.~\eqref{maincondition}, noting that we are allowed to
use this expression on the domain $\F_{n+2}$: For example, to apply the associativity
condition to the last two terms in the above expression, we need that
$r_{k\, k+1}< r_{k+1\,k+2} < r_{k \, k+2}$ for all $k$, which holds on $\F_{n+2}$. It is this property of the
domains $\F_i$ that motivates our definition~\eqref{Fndef}. Applying
the associativity condition, we find that all terms
cancel, thus proving the lemma. \qed

By this lemma, we can define a cohomology ring associated with the differential
$b$ as
\ben
H^n(V; \C) := \frac{Z^n(V; \C)}{B^n(V; \C)} = \frac{
\{ {\rm ker} \, b : \Omega^n(V) \to \Omega^{n+1}(V)\}
}{
\{
{\rm ran} \, b: \Omega^{n-1}(V) \to \Omega^n(V)
\}
} \, .
\een
As we will now see, the problem of finding a 1-parameter family of
perturbations $\C(x_1, x_2; \lambda)$ such that our associativity
condition~\eqref{maincondition} continues to
hold for $\C(x_1, x_2; \lambda)$ to all orders in $\lambda$ can
be elegantly and compactly be formulated in terms of this ring.
If we let
\ben
\C_i(x_1, x_2) = \frac{1}{i!} \, \frac{d^i}{d\lambda^i} \C(x_1, x_2; \lambda) \Bigg|_{\lambda = 0} \, ,
\een
then we note that the first order associativity condition,
\bena\label{maincondition1}
&&\C_0(x_2, x_3)\Big(\C_1(x_1, x_2) \otimes \id \Big) - \C_0(x_1, x_3)\Big(
\id \otimes \C_1(x_2, x_3) \Big) + \non\\
&&\C_1(x_2, x_3)\Big(\C_0(x_1, x_2) \otimes \id \Big) - \C_1(x_1, x_3)\Big(
\id \otimes \C_0(x_2, x_3) \Big) = 0\,\, ,
\eena
valid for $(x_1, x_2, x_3) \in \F_3$, is equivalent to the statement that
\ben
b \C_1 = 0 \, ,
\een
where here and in the following, $b$ is defined in terms of the unperturbed OPE-coefficient $\C_0$.
Thus, $\C_1$ has to be an element of $Z^2(V; \C_0)$. Let $z(\lambda): V \to V$
be a $\lambda$-dependent field redefinition in the sense of defn.~\ref{fieldred}, and
suppose that $\C(x_1, x_2)$ and $\C(x_1, x_2; \lambda)$ are connected
by the field redefinition. To first order, this means that
\ben\label{ctrivial}
\C_1(x_1, x_2) = -z_1 \C_0(x_1, x_2) + \C_0(x_1, x_2)(z_1 \otimes id + id \otimes z_1) \, ,
\een
or equivalently, that $bz_1 = \C_1$,
where $z_i = \frac{1}{i!} \, \frac{d^i}{d\lambda^i} z(\lambda) |_{\lambda=0}$.
Thus, the first order deformations of $\C_0$ modulo the trivial ones defined by
eq.~\eqref{ctrivial}
are given by the classes in $H^2(V; \C_0)$. The associativity condition for $i$-th order
perturbation (assuming that all perturbations up to order $i-1$ exist) can be written as
the following condition for $(x_1, x_2, x_3) \in \F_3$:
\bena\label{bciwi}
&& \C_0(x_2, x_3)\Big( \C_j(x_1, x_2) \otimes \id \Big)
- \C_j(x_1, x_3)\Big(
\id \otimes \C_0(x_2, x_3) \Big) +\\
&& \C_j(x_2, x_3)\Big( \C_0(x_1, x_2) \otimes \id \Big) - \C_0(x_1, x_3)\Big(
\id \otimes \C_j(x_2, x_3)\Big) = w_i(x_1, x_2, x_3) \non \, ,
\eena
where $w_i \in \Omega^3(V)$ is defined by
\ben
w_i(x_1, x_2, x_3) :=
-\sum_{j=1}^{i-1} \C_{i-j}(x_1, x_3)( \id \otimes \C_j(x_2, x_3) ) -
                  \C_{i-j}(x_2, x_3)( \C_j(x_1, x_2) \otimes \id ) \, .
\een
We assume here that all infinite sums implicit in this expression
converge on $\F_3$. This equation may be written alternatively as
\ben\label{bciwiup}
b \C_i = w_i \, .
\een
We would like to define the $i$-th order perturbation by solving
this linear equation for $\C_i$. Clearly, a necessary condition for there to
exist a solution is that $b w_i = 0$ or $w_i \in Z^3(V, \C_0)$, and this can indeed shown to
be the case, see lemma~\ref{obstrlemma} below. If a solution to eq.~\eqref{bciwiup}
exists, i.e. if $w_i \in B^3(V, \C_0)$, then any other solution will differ from
this one by a solution to the corresponding "homogeneous" equation.
Trivial solutions to the homogeneous equation of the form
$b z_i$ again correspond to an $i$-th order field redefinition and are not
counted as genuine perturbations. In summary, the perturbation series can be continued at $i$-th order
if $[w_i]$ is the trivial class in $H^3(V; \C_0)$,
so $[w_i]$ represents a potential $i$-th
order obstruction to continue the perturbation series.
If there is no obstruction, then the space of non-trivial
$i$-th order perturbations is given by $H^2(V; \C_0)$.
In particular, if we knew e.g. that $H^2(V; \C_0) \neq 0$ while
$H^3(V; \C_0) = 0$, then perturbations could be defined to arbitrary orders in
$\lambda$.

\begin{lemma}\label{obstrlemma}
If $w_i$ is in the domain of $b$, and if $b \C_j = w_j$ for all $j<i$, then $bw_i = 0$.
\end{lemma}

\medskip
\noindent
{\em Proof:} We proceed by induction in $i$. For $i=1$, the lemma
is true as we have
$w_1 = b \C_1$, so $bw_1 = 0$ by $b^2 = 0$. In the general case, using the
definition of $b$, we obtain the
following expression for $bw_i$:
\bena
&&-bw_i(x_1, x_2, x_3, x_4) \\
&&
=\sum_{j=1}^{i-1} \C_0(x_1, x_4)\Big( id \otimes \C_j(x_2, x_4)(id \otimes \C_{i-j}(x_3, x_4))\Big)\non\\
&&
-\sum_{j=1}^{i-1} \C_j(x_2, x_4)\Big( id \otimes \C_{i-j}(x_3, x_4) \Big) \Big( \C_0(x_1, x_2) \otimes id^2 \Big) \non\\
&&
+\sum_{j=1}^{i-1} \C_j(x_1, x_4)\Big( id \otimes \C_{i-j}(x_3, x_4)\Big)\Big( id \otimes \C_0(x_2, x_3) \otimes id \Big) \non\\
&&
-\sum_{j=1}^{i-1} \C_j(x_1, x_4)\Big( id \otimes \C_{i-j}(x_2, x_4) \Big)\Big( id^2 \otimes \C_0(x_3, x_4) \Big) \non\\
&&
+\sum_{j=1}^{i-1} \C_0(x_3, x_4)\Big( \C_j(x_1, x_3)(id \otimes \C_{i-j}(x_2, x_3)) \otimes id \Big) \non\\
&&
-\sum_{j=1}^{i-1} \C_0(x_1, x_4)\Big( id \otimes \C_j(x_3, x_4)(\C_{i-j}(x_2, x_3) \otimes id) \Big) \non\\
&&
+\sum_{j=1}^{i-1} \C_j(x_3, x_4)\Big( \C_{i-j}(x_2, x_3) \otimes id \Big) \Big( \C_0(x_1, x_2) \otimes id^2\Big) \non\\
&&
-\sum_{j=1}^{i-1} \C_j(x_3, x_4)\Big( \C_{i-j}(x_1, x_3) \otimes id \Big) \Big( id \otimes \C_0(x_2, x_3) \otimes id \Big) \non\\
&&
+\sum_{j=1}^{i-1} \C_j(x_2, x_4)\Big( \C_{i-j}(x_1, x_2) \otimes id\Big)  \Big( id^2 \otimes \C_0(x_3, x_4) \Big) \non\\
&&
-\sum_{j=1}^{i-1} \C_0(x_3, x_4)\Big( \C_j(x_2, x_3)(\C_{i-j}(x_1, x_2) \otimes id) \otimes id \Big) \non \, .
\eena
After some manipulations using the definition of $b$ and that by definition the points
$(x_1, x_2, x_3, x_4)$ are assumed to be in ${\mathcal F}_4$,
we can transform this into the following expression
\bena
&& -bw_i(x_1, x_2, x_3, x_4) \\
&=&+\sum_{j=1}^{i-1} b \C_j(x_1, x_2, x_4)(id^2 \otimes \C_{i-j}(x_3,x_4)) \non\\
&&-\sum_{j=1}^{i-1} \C_j(x_1, x_4)(id \otimes b\C_{i-j}(x_2, x_3, x_4))\non\\
&&-\sum_{j=1}^{i-1}b\C_j(x_1, x_3, x_4)(id \otimes \C_{i-j}(x_2, x_3) \otimes id)\non\\
&&-\sum_{j=1}^{i-1} \C_j(x_3, x_4)(b\C_{i-j}(x_1, x_2, x_3) \otimes id) \non\\
&&+\sum_{j=1}^{i-1} b\C_j(x_2, x_3, x_4)(\C_{i-j}(x_1, x_2) \otimes id^2)  \, ,\non
\eena
where the first sum comes from the first two sums of the previous
equation, the second from the third and fourth two sums, etc.
We now substitute the relation $b \C_j = w_j$ for $j\le i-1$ on ${\mathcal F}_3$, noting
that we are allowed to do so when $(x_1, x_2, x_3, x_4) \in {\mathcal F}_4$: For example,
in the last term $(x_2, x_3, x_4) \in {\mathcal F}_3$ is satisfied whenever
$(x_1, x_2, x_3, x_4) \in {\mathcal F}_4$, and a similar statement holds for the
other 4 terms [this is in fact our motivation for our definition of the domains $\F_n$].
We then perform the sum over $j$. If this is done, then we see that the five terms in the sum become ten terms
involving each three factors of the $\C$'s. These terms cancel pairwise, and
we get the desired result that $bw_i=0$, as we desired to show. \qed

\section{Gauge Theories}\label{hochschild}

Local gauge theories are typically more complicated than
theories without local gauge invariance. One way to understand the
complicating effects due to local gauge invariance is to realize that
the dynamical field equations are not hyperbolic in nature in Lorentzian
spacetimes. This is seen most clearly
in the case of classical field theories. Because local gauge transformations may
be used to change the gauge connection in arbitrary compact regions of spacetime, it is
clear that the gauge connection cannot be entirely determined by the dynamical
equations and its initial data on some spatial time slice.
Thus, there is no well-posed initial value formulation in the standard sense. Similar
remarks apply to the Euclidean situation.

To circumvent this problem, one typically proceeds in two steps. At the first step, an
auxiliary theory is considered, containing the gauge fields as well as additional
"ghost" fields taking values in an infinite-dimensional Grassmann algebra.
This theory has a well-posed initial value formulation. At the second
step, the new degrees of freedom are removed. Here it is important that the auxiliary
theory possesses a new symmetry, the so-called BRST-symmetry, $s$, which is a linear
transformation on the space of classical fields with the property $s^2=0$ [for example,
in Yang-Mills theory $s$ is given by eq.~\eqref{BRSTt}]. It turns out
that the field content and dynamics
of the original theory may be recovered by considering only
the equivalence classes of fields in the auxiliary theory that are in the
null-space of $s$, modulo those that are in the range of $s$. Thus, the second step is
to define the observables of the gauge theory in question as the {\em cohomology}
of the "differential" $s$.

At the quantum level, one has a similar structure. In the framework considered in this
paper, the situation may be described abstractly as follows: As before, we have an
abstract vector space of fields, $V$. This space is to be thought of as the collection
of the components of all (composite) fields in the {\em auxiliary} theory including
ghost fields. The space $V$ is equipped with a grading $\gamma$ and a differential
$s$, i.e., two linear maps
\ben
s: V \to V \, , \quad \gamma: V \to V \, ,
\een
with the properties
\ben\label{sprop}
s^2 = 0 \, , \quad \gamma^2 = id \, \quad \gamma s + s \gamma = 0 \, .
\een
The map $s$ should be thought of as being analogous to the classical BRST-transformation.
The map $\gamma$ has eigenvalues $\pm 1$, and the eigenvectors
correspond as above to Bose/Fermi fields. At the classical level, the elements in the eigenspace of $-1$
are analogous to the classical (composite) fields of odd Grassmann parity, while those
in the eigenspace of $+1$ are analogous to those of even Grassmann parity. However,
we emphasize that these are just analogies, as we will be dealing with a quantum field
theory. For the general analysis of quantum gauge theories
we will only need $s$ and $\gamma$ to satisfy the
above properties. It is also natural to postulate the existence of another grading map
$g: V \to V$ with the properties ${\rm Spec} \, g = \mathbb Z$ and
$sg = (g + id) s$, $\gamma g - g \gamma = 0$.
This map is to be thought of as the number counter for the ghost fields (so that
$s$ increases the ghost number by one unit). Finally, we would like all maps
$s, \gamma, g$ to be compatible with the $\star$-operation on $V$, and to
preserve the grading by the spin, as well as the dimension.

We next consider a quantum field theory whose fields are described by the elements
of $V$, with operator product coefficients $\C$. At the classical level, $s$ is
a graded derivation, so we would also like $s$ to be a graded derivation at the quantum
level. Recall that if $\A$ is a graded algebra with grading map $\Gamma$ (i.e.,
$\Gamma^2=id$), then a graded derivation
is a map $D: \A \to \A$ with the property that
\ben\label{DAB}
D(AB) = (DA) B + \Gamma(A) DB \quad \text{for all $A,B \in \A$} \, .
\een
Equivalently, if we write the product in the algebra as $m: \A \otimes \A \to \A$
with $m(A,B) = AB$, then $m$ should satisfy
\ben
D m = m(D \otimes id) + m(\Gamma \otimes D) \, ,
\een
in the sense of maps $\A \otimes \A \to \A$.
As we have emphasized several times, the OPE-coefficients $\C(x_1, x_2)$
are to be thought of informally as the expansion coefficients of
a product. Therefore, if $s$ is to be a graded derivation we should
add a corresponding additional axiom to those formulated above in sec.~\ref{axiomatic}.
Heuristically, we want $s$ to act on a product of quantum fields $\phi_a$
in the following way analogous to eq.~\eqref{DAB}:
\ben
s\Big[ \prod_{i=1}^n \phi_{a_i}(x_i) \Big] =
\sum_{i=1}^n (-1)^{\sum_{j<i} \epsilon_i} \phi_{a_1}(x_1) \cdots
s \phi_{a_i}(x_i) \cdots \phi_{a_n}(x_n) \, ,
\een
Here, $\epsilon_i = 0,1$ according to whether $\phi_{a_i}$ is
bosonic or fermionic. If we formally apply an OPE to both sides of this equation, then we
arrive at the following condition for the OPE coefficients:

\medskip
\noindent
\paragraph{\bf BRST-invariance:} The OPE coefficients of the auxiliary
should satisfy the additional condition
\ben\label{BRSTn}
s \C(x_1, \dots, x_n) = \sum_{i=1}^n \C(x_1, \dots, x_n)(\gamma^{i-1} \otimes
s \otimes id^{n-i})
\een
for all $n$.

\medskip
\noindent
Above, we have seen in prop.~\ref{proposition1} that the 2-point OPE coefficients determine
all higher coefficients uniquely. Thus, as a corollary, the above conditions of
$BRST$-invariance will be satisfied if they hold for the 2-point coefficients,
i.e. if the condition
\ben\label{scompat}
s \C(x_1, x_2) = \C(x_1, x_2) (s \otimes id) + \C(x_1, x_2)(\gamma \otimes s)
\een
holds.
Furthermore,
we would like to formulate abstractly the condition that, since the OPE coefficients
are valued in the complex numbers, they should have "ghost number" equal to zero,
meaning that
\ben\label{gcompat}
g \C(x_1, x_2) = \C(x_1, x_2)(g \otimes id) + \C(x_1, x_2)(id \otimes g) \, .
\een
In summary a quantum gauge theory is described in our language abstractly as follows:

\begin{defn}
A quantum gauge theory is a system of OPE-coefficients
\ben
\C = (\C(x_1, x_2), \C(x_1,x_2,x_3), \dots)
\een
associated with $V$ satisfying the properties laid out in sec.~\ref{axiomatic},
together with a ghost number grading $g$ satisfying~\eqref{gcompat},
and a differential $s:V \to V$ satisfying~\eqref{scompat} and~\eqref{sprop}, as
well as $(g + id) s = sg$.
\end{defn}

By analogy with the classical case, we define the space of {\em physical fields}
of the gauge theory to be the quotient
\ben
\widehat V := \frac{\{{\rm ker} \, s: V^0 \to V^{+1} \}}{\{{\rm ran} \, s: V^{-1} \to V^0 \}}
\een
where $V^q$ are the
eigenspaces of the linear map $g$, with eigenvalue $q$,
\ben
V = \bigoplus_{q \in \mathbb Z} V^q \, , \quad s: V^q \to V^{q+1} \, .
\een
In other words, we define the space of physical fields as the zeroth cohomology group
defined by $s$, with the general cohomology group at $q$-th order defined by
\ben
H^q(V;s) = \frac{\{\ker s: V^q \to V^{q+1}\}}{\{\ran s: V^{q-1} \to V^q \}} \, .
\een
Because the OPE coefficients satisfy the assumption of BRST~invariance, eq~\eqref{BRSTn},
we have the following proposition/definition:

\begin{prop}\label{factorprop}
The OPE coefficients $\C$ of the auxiliary theory induce maps
\ben
\widehat \C(x_1, \dots, x_n): \widehat V \otimes \dots \otimes \widehat V \to \widehat V \, ,
\een
so the operator product expansion "closes" on the space $\widehat V$ of physical fields.
Therefore, the {\em true physical sector of the gauge theory} can be defined as
the quantum field theory described by the pair $(\widehat V, \widehat \C)$.
\end{prop}
\noindent
\paragraph{\bf Remarks:} 1) In Yang-Mills theory with Lie algebra $\frak g$, the space $V$ is naturally identified with the free
unital commutative $\partial_\mu$-differential module (over $\mc[\lambda]$) generated by the formal
expressions of the form $\myid$ and
\ben
\partial_{\mu_1} \dots \partial_{\mu_k} \psi_i ; \,\, \mu_j = 1, \dots, D  \, ,
\een
where $\psi_i$ denotes either a component of $A$ or the auxiliary "field" $F$ or the ghost "fields", $U,\bar U$.
The expressions in
$V$ are taken modulo the relations $\psi_i \psi_j = (-1)^{F_i F_j} \psi_j \psi_i$, with
$F_i = 0$ or $=1$ according to whether $g(\psi_i) = \pm \psi_i$, where $g$ is $-1$ on
ghost fields $U,\bar U$, and $+1$ on $A,F$. Furthermore,
on $V$, the linear maps $\partial_\mu$ are defined to act as the (ungraded) derivations that are obtained by
formally viewing the elements of $V$ as classical fields. On $V$, there also acts the BRST-differential
$s$. It is defined to act on the generators of $V$ by eq.~\eqref{BRSTt}, and it is demanded to anti-commute
with the formal derivations $\partial_\mu$, i.e.,
\ben
\partial_\mu \in {\rm Der}(V) \, , \quad s \circ \partial_\mu = \partial_\mu \circ s \, , \quad
g \circ \partial_\mu = \partial_\mu \circ g \, .
\een
One can then show~\cite{barnich} that $\widehat V$
corresponds precisely to the gauge-invariant monomials of the
field strength tensor of the gauge connection and its covariant derivatives, i.e.,
\ben
\widehat V = \Big\langle p(\D^{k_1} F, \dots, \D^{k_n}F) ; p \in {\rm Inv}({\mathfrak g}^{\otimes n}, \mc) \Big\rangle \, ,
\een
where ${\rm Inv}(\mathfrak g^{\otimes n}, \mc)$ is the space of $\frak{g}$-invariant multi-linear forms on Lie-algebra,
$\D_\mu = \partial_\mu + i\lambda[A_\mu, \, . \, ]$ is the
standard covariant derivative, $F$ is a shorthand for its curvature, $F_{\mu\nu} = [\mathcal{D}_\mu, \mathcal{D}_\nu]$,
and $\D^k$ is a shorthand for $\D_{(\mu_1} \cdots \D_{\mu_k)}$.

\medskip
\noindent
2) Note that the OPE-coefficients of the auxiliary theory
not only close on the space $\widehat V$, but more generally on any of the spaces
$W_k = \oplus_{q \ge k} H^q(V; s)$. These spaces contain also operators of non-zero
ghost number. One does not, however, expect this theory to have any non-trivial
states satisfying the OS-positivity axiom [see sec.~\ref{axiomatic}].

\medskip
\noindent
{\em Proof}: Let $|v_1 \rangle, \dots, |v_n\rangle \in {\rm ker} \, s$. Using eq.~\eqref{BRSTn}, we have
\bena
&&s \Big( \C(x_1, \dots, x_n)|v_1 \otimes \dots \otimes v_n \rangle \Big) \\
&=& \sum_{i=1}^n \C(x_1, \dots, x_n) | \gamma(v_1) \otimes \dots \gamma(v_{i-1}) \otimes sv_i \otimes v_{i+1} \otimes \dots v_n \rangle = 0 \, . \non
\eena
Thus, the composition $\C(x_1, \dots, x_n)|v_1 \otimes \dots \otimes v_n \rangle$ is in the kernel of $s$. One similarly shows that
if $|v_1 \rangle, \dots, |v_n\rangle \in {\rm ker} \, s$, and in addition $|v_i \rangle \in {\rm ran} \, s$ for some $i$, then
the composition is even in the image of $s$. Thus, $\C(x_1, \dots, x_n)$ gives a well defined map
from $({\rm ker} \, s/{\rm ran} \, s)^{\otimes n}$ into ${\rm ker} \, s/{\rm ran} \, s$.
Finally, since $\C(x_1, \dots, x_n)$ satisfies the analogue of eq.~\eqref{gcompat}, it follows that
the composition has ghost number zero if each $|v_i\rangle$ has. Thus, $\C(x_1, \dots, x_n)$ gives a well defined map
$\widehat \C(x_1, \dots, x_n)$
from $\widehat V^{\otimes n}$ to $\widehat V$. This map inherits the properties of
factorization, scaling, the unity axiom, the symmetry property etc. from the map $\C(x_1, \dots, x_n)$.
Thus, the collection $(\widehat \C, \widehat V)$ again defines a quantum field theory in our sense.
\qed

\medskip

We would now like to consider perturbations of a given quantum gauge theory
by analogy with the procedure described in the previous section.
Thus, as above, let $\lambda$ be a formal expansion parameter, and let
$\C(x_1, x_2; \lambda)$ be a 1-parameter family describing a deformation of
the given 2-point OPE coefficient of the auxiliary theory.
As above let $(\C_0, \C_1, \C_2, \dots)$ be the zeroth, first, second,
etc. perturbations of the expansion
coefficients. In order that the perturbed coefficients satisfy the
associativity constraint, the equations~\eqref{bciwi} must again hold for the coefficients.
In the situation at hand, we also should consider a deformation $s(\lambda)$ of
the BRST-differential, with expansion coefficients $(s_0, s_1, s_2, \dots)$,
\ben
s_i = \frac{1}{i!} \, \frac{d^i}{d\lambda^i} \, s(\lambda) \Bigg|_{\lambda = 0} \, .
\een
These quantities should satisfy the perturbative version of eq.~\eqref{sprop}, that is
\ben\label{sconsi}
\sum_{j=0}^i s_j s_{i-j} = 0 \, , \quad s_i \gamma + \gamma s_i = 0 \, ,
\een
and they should satisfy the perturbative version of eq.~\eqref{scompat},
\ben\label{cconsi}
\sum_{j=0}^i s_j \C_{i-j}(x_1, x_2) = \sum_{j=0}^i \C_{i-j}(x_1, x_2) (s_j \otimes id)
+ \sum_{j=0}^i \C_{i-j}(x_1, x_2)(\gamma \otimes s_j) \, ,
\een
for all $i = 0, 1, 2, \dots$. For $i=0$, these conditions are just the
conditions that the undeformed theory described by $s_0, \C_0$ defines a gauge theory. For $i=1, 2, \dots$,
we get a set of conditions that constrain the possible $i$-th order perturbations
$s_i, \C_i$. Actually, as in the previous section,
we would like to exclude again that our deformations $s_i, \C_i$ are
simply due to a $\lambda$-dependent field redefinition, see defn.~\ref{fieldred}. In the present context,
a first order perturbation $s_1, \C_1$ that is simply due to a field redefinition is
one for which
\ben\label{ctrivial1}
\C_1(x_1, x_2) = -z_1 \, \C_0(x_1, x_2) + \C_0(x_1, x_2) (z_1 \otimes id + id \otimes z_1)
  \, , \quad s_1 = s_0 \, z_1 + z_1 \, s_0 \, ,
\een
for some $z_1: V \to V$ such that $z_1 \, \gamma = \gamma \, z_1$. There are
similar conditions at higher order. We will now see that
the higher order conditions, have an elegant formulation
in terms of a variant of Hochschild cohomology associated with $\C_0$,
twisted with the cohomology of $s_0$.

In order to describe this, we begin by defining the respective cohomology rings.
Our first task is the definition of the Hochschild type differential
$b$ in the case when $V$ is a graded vector space. Let $\C(x_1, x_2): V \otimes V \to V$
satisfy the associativity condition~\eqref{maincondition} and be even under our grading
$\gamma$, meaning $\C(x_1, x_2)(\gamma \otimes \gamma) = \gamma \C(x_1, x_2)$.

\begin{defn}
Let $\Omega^n(V)$ be the space of all translation invariant analytic maps
$f_n: \F_{n} \to \hom(V \otimes \dots \otimes V, V)$, where $\F_n \subset
(\mr^D)^n$ is the domain~\eqref{Fndef}. Let
\ben
f^\gamma_n := \gamma f_n (\gamma \otimes \dots \otimes \gamma) \, .
\een
If $f_n^\gamma = f_n$, then $f_n$ is said to be even and
the definition of $b f_n \in \Omega^{n+1}(V)$ is
as above in eq.~\eqref{bfndef}. If $f_n^\gamma = -f_n$, then $f_n$ is
said to be odd, and we define
\bena\label{bdefodd}
(b f_n)(x_1, \dots, x_{n+1}) &:=& -\C(x_1, x_{n+1})(\gamma \otimes f_n(x_2, \dots, x_{n+1}) \non\\
&-& \sum_{i=1}^n (-1)^i f_n( x_1, \dots, \widehat x_i, \dots, x_{n+1})
(id^{i-1} \otimes \C(x_i, x_{i+1}) \otimes id^{n-i}) \non\\
&-& (-1)^{n+1} \C(x_n, x_{n+1}) (f_n(x_1, \dots, x_n) \otimes id) \, .
\eena
\end{defn}

As in the definition of $b$ in the ungraded case, we may check that
$b^2 = 0$, so we may again define the cohomology of $b$ as above. We next prove
a simple lemma about the relation between the differential $b$ and the differential
$s$ when the quantum field theory is a gauge theory $(V, \C, s)$. First,
we define an action of $s$ on the space $\Omega^n(V)$ of analytic maps $f_n$
by $B: \Omega^n(V) \to \Omega^n(V)$, where
\bena\label{deltadef}
(B f_n)(x_1, \dots, x_n) &:=& s f_n(x_1, \dots, x_n) \non\\
&-& \sum_{i=1}^n f^\gamma_n(x_1, \dots, x_n)(\gamma^{i-1} \otimes s \otimes id^{n-i}) \, .
\eena

\begin{lemma}
We have $B (B f_n) =0$ for all $f_n$.
If $f_n$ is in the domain of $b$, then $B b f_n = - b B f_n$. Symbolically
\ben
b B + B b = 0, \quad B^2 = 0 \, .
\een
\end{lemma}

\noindent
{\em Proof:} For the proof of the first statement we consider first
the case when $f_n^\gamma = f_n$, and we apply $B$ one more time
to eq.~\eqref{deltadef}. We obtain the following three terms:
\bena
&&B (B f_n)(x_1, \dots, x_n) = s^2 f_n(x_1, \dots, x_n) \\
&&-\sum_{i=1}^n [(sf_n)^\gamma(x_1, \dots, x_n)+sf_n^\gamma(x_1, \dots, x_n)](\gamma^{i-1} \otimes s \otimes id^{n-i})\non\\
&&+\sum_{i,j=1}^n  f_n(x_1, \dots, x_n)(\gamma^{i-1} \otimes s \otimes id^{n-i})
(\gamma^{j-1} \otimes s \otimes id^{n-j}) \non
\, .
\eena
The first term vanishes since $s^2=0$. The second term vanishes because if
$f_n$ is even under $\gamma$, then $sf_n$ is odd, so $(sf_n)^\gamma + sf_n^\gamma = 0$.
We split the double sum into three parts---the terms for which $i<j$, the terms for
$i>j$, and the terms for which $i=j$. The third set of terms give zero using $s^2 = 0$.
The first set of terms is manipulated using $s \gamma = - \gamma s$:
\bena
&&+\sum_{i<j}  f_n(x_1, \dots, x_n)
(\gamma^{i-1} \otimes s \otimes id^{n-i})
(\gamma^{j-1} \otimes s \otimes id^{n-j})
\non\\
&=&\sum_{i<j}  f_n(x_1, \dots, x_n)(id^{i-1} \otimes s\gamma \otimes \gamma^{j-i-1}
\otimes s \otimes id^{n-j})
\non\\
&=&-\sum_{i<j}  f_n(x_1, \dots, x_n) (\gamma^{j-1} \otimes s \otimes id^{n-j})
(\gamma^{i-1} \otimes s \otimes id^{n-i})
\, .
\eena
After changing the names of the indices, this is seen to be
equal to minus the second set of terms, so $B (B f_n) = 0$.
The case $f_n^\gamma = -f_n$ is completely analogous.

We next prove the relation $b(B f_n) = - B (b f_n)$, again
assuming for definiteness that $f^\gamma_n = f_n$. To compute
$b(B f_n)$,  we apply $b$ to eq.~\eqref{deltadef}, and
use that $(B f_n)^\gamma = - B f_n$. This gives
\bena\label{bdel}
&-& b(B f_n)(x_1, \dots, x_{n+1}) = \C(x_1, x_{n+1})[\gamma \otimes sf_n(x_2, \dots, x_{n+1})] \non\\
&-& \sum_{i=1}^n \C(x_1, x_{n+1})[\gamma \otimes f_n(x_2, \dots, x_{n+1})
(\gamma^{i-1} \otimes s \otimes id^{n-i})] \non\\
&+& \sum_{i=1}^n (-1)^i sf_n(x_1, \dots, \widehat x_i, \dots, x_n)(id^{i-1} \otimes \C(x_i, x_{i+1}) \otimes id^{n-i}) \non\\
&-& \sum_{i,j=1}^n (-1)^i f_n(x_1, \dots, \widehat x_i, \dots, x_n)(\gamma^{j-1} \otimes s \otimes id^{n-j}) (id^{i-1} \otimes \C(x_i, x_{i+1}) \otimes id^{n-i}) \non\\
&+& (-1)^{n+1} \C(x_n, x_{n+1}) [s f_n(x_1, \dots, x_n) \otimes id] \non\\
&-& (-1)^{n+1}
\sum_{i=1}^n \C(x_n, x_{n+1}) (f_n(x_1, \dots, x_n) \otimes id)(\gamma^{i-1} \otimes s \otimes id^{n-i+1}) \, .
\eena
We next evaluate $B(b f_n)$ by applying $B$ to eq.~\eqref{bfndef}. This gives
\bena\label{delb}
&& B(b f_n)(x_1, \dots, x_{n+1}) = s \C(x_1, x_{n+1}) [id \otimes f_n(x_2, \otimes, x_{n+1})]\non\\
&+& \sum_{i=1}^n (-1)^i \, s f_n(x_1, \dots, \widehat x_i, \dots, x_n)
[\id^{i-1} \otimes \C(x_i, x_{i+1}) \otimes id^{n-i}] \non\\
&+& (-1)^{n+1} \, s \C(x_n, x_{n+1})[f_n(x_1, \dots, x_n) \otimes id] \non\\
&-& \sum_{i=1}^{n+1}  \C(x_1, x_{n+1}) [id \otimes f_n(x_2, \otimes, x_{n+1})](\gamma^{i-1} \otimes s \otimes id^{n+1-i}) \non\\
&-& \sum_{j=1}^{n+1} \sum_{i=1}^n (-1)^i \, f_n(x_1, \dots, \widehat x_i, \dots, x_n)
[\id^{i-1} \otimes \C(x_i, x_{i+1}) \otimes id^{n-i}](\gamma^{j-1} \otimes s \otimes id^{n+1-j}) \non\\
&-& (-1)^{n+1} \sum_{i=1}^{n+1} \C(x_n, x_{n+1})[f_n(x_1, \dots, x_n) \otimes id] (\gamma^{i-1} \otimes
s \otimes id^{n+1-i}) \, .
\eena
We next bring $s$ behind $\C$ in all terms in this expression using eq.~\eqref{scompat},
and we use that $\C$ itself is even under $\gamma$. If these
steps are carried out, then it is seen that all terms in eq.~\eqref{bdel} match
a corresponding term in eq.~\eqref{delb}. The calculation when $f_n^\gamma = -f_n$
is again analogous.
\qed

The fact that $b^2 = 0$ and the
properties of $B$ and $b$ stated in the lemma imply that $(B+b)^2 = B^2 + b^2 + bB + Bb = 0$.
Hence the map
\ben
\delta := B + b \,, \quad \quad \delta : \bigoplus_n \Omega^n(V) \to \bigoplus_n \Omega^n(V)
\een
is again a differential, i.e., it satisfies $\delta^2 = 0$. Therefore, we can
again define a corresponding cohomology ring
\ben
H^n(\delta; V) := \frac{\{(f_1, f_2, \dots, f_n, 0, 0, \dots) \in \ker \delta\}}{
\{(f_1, f_2, \dots, f_n, 0, 0, \dots) \in \ran \delta \}} \equiv
\frac{Z^n(\delta; V)}{B^n(\delta; V)} \, .
\een
Thus, a general element in $H^n(\delta; V)$ consists of an equivalence class
of a sequence
\ben
(f_1, f_2, \dots, f_n, 0, 0, \dots)\,, \quad Bf_1 = bf_n = 0 \, , \quad bf_{i-1} = -Bf_i \, \quad \text{for $1<i\le n$} \, ,
\een
where each $f_i$ is an element in $\Omega^i(V)$ and $n$ is some finite number,
modulo all sequences with the property that there exist $h_i \in \Omega^i(V) \cap {\rm dom} \, b$
for $1 \le i < n$ such that
\ben
(f_1, f_2, \dots, f_n, 0, 0, \dots)\,, \quad f_1 = Bh_1\, , \quad f_n = bh_{n-1} \,, \quad f_i = bh_{i-1} + Bh_i \, ,
\een
for all $1<i<n$. The conditions~\eqref{sconsi}, \eqref{cconsi}, \eqref{bciwi} expressing
respectively the nilpotency of the perturbed BRST operator
$s_i$, the compatibility of the BRST operator with the perturbations
$\C_i$ of the operator product, and the corresponding associativity condition at the $i$-th
order in perturbation theory may now be expressed by a simple condition
in terms of this cohomology ring. For this, we define the
differentials $b, B$ and $\delta = B+b$ as
above in terms of the unperturbed theory, i.e. using
$\C_0$ and $s_0$. For $i>0$, we combine
$s_i$ and $\C_i$ into the element
\ben
\beta_i := (s_i, \C_i, 0, 0, \dots) \in \bigoplus_n \Omega^n(V) \, .
\een
and we define $\alpha_i = (u_i, v_i, w_i, 0, 0, \dots)$, where
\bena
u_i(x_1) &:=& -\sum_{j=1}^{i-1} s_j s_{i-j} \,, \\
v_i(x_1, x_2) &:=& -\sum_{j=1}^{i-1} s_j \C_{i-j}(x_1, x_2) -
\C_{i-j}(x_1, x_2) (s_j \otimes id) - \C_{i-j}(x_1, x_2)(\gamma \otimes s_j) \, , \non\\
w_i(x_1, x_2, x_3) &:=& -\sum_{j=1}^{i-1} \C_j(x_1, x_3)[id \otimes \C_{i-j}(x_2, x_3)]
- \C_j(x_2, x_3)[\C_{i-j}(x_1, x_2) \otimes id] \,  . \non
\eena
The conditions~\eqref{sconsi}, \eqref{cconsi}, \eqref{bciwi} can now be simply
and elegantly be restated as the single condition
\ben\label{betacond}
\delta \beta_i = \alpha_i \, .
\een
This is the desired cohomological formulation of our consistency
conditions for perturbations of a gauge theory.

Let us analyze the conditions~\eqref{betacond} on $\beta_i$. First we
note that $\alpha_1 = 0$, and that $\alpha_i$ is defined in terms
of $\beta_1, \beta_2, \dots, \beta_{i-1}$ for $i>1$.
When $i=1$, the above condition hence states that $\delta \beta_1 = 0$,
meaning that $\beta_1 \in Z^2(\delta; V)$. On the other hand, we can
express the situation when
$s_1$ and $\C_1$ merely correspond to a
field redefinition [see eq.~\eqref{ctrivial1}] as saying that
\ben
\beta_1 = \delta \zeta_1 \, ,
\een
where $\zeta_1 \equiv (z_1, 0, 0, \dots)$ is given in terms of the first order
field redefinition $z_1$. Thus, in this case $\beta_1 \in B^2(\delta; V)$.
In summary, the first order perturbations of the BRST-operator and of the product modulo
the trivial ones are in one-to-one correspondence with the non-trivial elements
of the ring $H^2(V; \delta)$. Let us now assume that we have picked a non-trivial
first order perturbation $\beta_1$---assuming that such a perturbation exists. Then
$\beta_2$ must satisfy eq.~\eqref{betacond}, $\delta \beta_2 = \alpha_2$, for
the $\alpha_2$ calculated from $\beta_1$. Clearly, because $\delta^2 = 0$,
a necessary condition for the existence of a solution to eq.~\eqref{betacond}
is that $\delta \alpha_2 = 0$, meaning that $\alpha_2 \in Z^3(\delta; V)$. This
can indeed be checked to be the case (see the lemma below).
Our requirement that $\delta \beta_2 = \alpha_2$ is however a stronger
statement, meaning that in fact $\alpha_2 \in B^3(V; \delta)$. Thus, if the class
$[\alpha_2]$ in $H^3(\delta; V)$ is non-trivial, then no second order perturbations
to our gauge theory exists, or said differently, $[\alpha_2] \in H^3(\delta; V)$
is an obstruction to continue the deformation process.

Let us assume that there is no obstruction so that
a solution $\beta_2$ to the "inhomogeneous equation" $\delta \beta_2 = \alpha_2$ exists.
Any solution to the equation will only be unique up to a solution to the corresponding
"homogeneous equation" $\delta \beta_2 = 0$. In fact, because any solution
to the inhomogeneous equation can be written as an arbitrary but fixed solution
plus the general solution to the homogeneous equation, it follows that
the second order perturbations $\beta_2$ are parametrized by the elements of
$Z^2(\delta; V)$. Special solutions to the homogeneous equation
include in particular ones of the form $\beta_2 = \delta \zeta_2 \in B^2(\delta; V)$,
with $\zeta_2 \equiv (z_2, 0, 0, \dots)$. However, any such solution of the
homogeneous equation can again be absorbed into a second order field redefinition
parametrized by $z_2$. Thus, we see that if the obstruction $[\alpha_2]$
vanishes at second order, then the second order perturbations modulo the trivial
perturbations are again parametrized by the elements of the space $H^2(\delta; V)$.

In the general order, we assume inductively that a solution to the
consistency relations $\delta \beta_j = \alpha_j$ has been found for all
$j<i$, meaning in particular that the obstructions
$[\alpha_j]$ vanish for all $j<i$.
By the lemma below, $\delta \alpha_i = 0$, so
$\alpha_i$ defines a class $[\alpha_i] \in H^3(\delta; V)$.
If this class if non-trivial, then the deformation process cannot be continued. If
it is the trivial class, by definition there is a solution $\beta_i$ to the
equation $\delta \beta_i = \alpha_i$. Again, this is unique only up to a solution to the
corresponding homogeneous equation $\delta \beta_i = 0$. The non-trivial solutions among these
not corresponding to a field redefinition are again in one-to-one correspondence with
the elements in the ring $H^2(\delta; V)$. Thus, a sufficient condition for
there to exist a consistent, non-trivial perturbation to the product and
BRST operator to arbitrary order in perturbation theory is
\ben
H^2(\delta; V) \neq 0\, , \quad H^3(\delta; V) = 0 \, ,
\een
for in this case all obstructions are trivial.
Moreover, in that case, $H^2(\delta; V)$ parameterizes all non-trivial $i$-order
perturbations for any $i \ge 1$.

\begin{lemma}
Assume that $\delta \beta_j = \alpha_j$ for all $j<i$, or equivalently,
that $[\alpha_j] \in H^3(\delta; V)$ defines the trivial element for
all $j<i$, and assume that the chain $\alpha_i$ is in the domain of $\delta$ for all $i$. Then we have
$\delta \alpha_i = 0$. In component form
\ben
B u_i = 0 \, , \quad bu_i + Bv_i = 0 \, , \quad bv_i + Bw_i = 0 \, ,
\quad bw_i = 0 \, .
\een
\end{lemma}

\noindent
{\em Proof:} For a given $i$, the hypothesis of the lemma amounts to
saying that $Bs_j = u_j, bs_j + B\C_j = v_j$ and $b\C_j = w_j$ for all
$j<i$. It follows from the last equation that $bw_i = 0$, as we
have already proved above in lemma~\ref{obstrlemma} above.

We next concentrate on proving the relation $Bu_i = 0$. We have
\ben\label{Bf1}
B u_i = -\sum_{j=1}^{i-1} (Bs_j) s_{i-j} + \sum_{j=1}^{i-1} s_{i-j} (Bs_j)  \, .
\een
Now, using that $Bs_j = u_j$ for the perturbations at order
$j<i$ and the definition of $u_j$, the first sum is equal to
\bena
&&\sum_{j=1}^{i-1} (Bs_j) s_{i-j} =\sum_{j=1}^{i-1} \sum_{k=1}^{j-1} s_k s_{j-k} s_{i-j}
\non\\
&=&\sum_{j=1}^{i-1} \sum_{k=1}^{i-j-1} s_j s_{i-j-k} s_k =
\sum_{j=1}^{i-1} s_{i-j} (Bs_j)
\, .
\eena
Thus, the first and second sum in~\eqref{Bf1} precisely cancel, and we have shown
$Bu_i = 0$.
We next show that $bu_i + Bv_i = 0$. A straightforward calculation using
the definitions of $v_i$ and of $B$ gives
\bena
&&Bv_i(x_1, x_2) = \\
&&\sum_{j=1}^{i-1} -(Bs_j) \Big( \C_{i-j}(x_1, x_2) \Big) + s_j \Big( B\C_{i-j}(x_1, x_2) \Big)      \non\\
&&\sum_{j=1}^{i-1} +\C_{i-j}(x_1, x_2) \Big( Bs_j \otimes id + id \otimes Bs_j \Big)       \non\\
&&\sum_{j=1}^{i-1} +B\C_{i-j}(x_1, x_2) \Big( s_j \otimes id + \gamma \otimes s_j \Big) \, .\non
\eena
By the assumptions of the lemmas, we may substitute
$B\C_j = v_j - bs_j$ and $Bs_j = u_j$ for $j<i$.
This leads to
\bena
&&Bv_i(x_1, x_2) = \\
&&\sum_{j=1}^{i-1} -u_j \Big( \C_{i-j}(x_1, x_2) \Big) - s_j \Big( bs_{i-j}(x_1, x_2) \Big)      \non\\
&&\sum_{j=1}^{i-1} +\C_{i-j}(x_1, x_2) \Big( u_j \otimes id + id \otimes u_j \Big)       \non\\
&&\sum_{j=1}^{i-1} -bs_{i-j}(x_1, x_2) \Big( s_j \otimes id + \gamma \otimes s_j \Big) \non\\
&&\sum_{j=1}^{i-1} +s_j v_{i-j}(x_1, x_2) + v_{i-j}(x_1, x_2)\Big( s_j \otimes id + \gamma \otimes s_j \Big)
\, .\non
\eena
We now use again the definition of $b$ and we substitute
the expressions for $v_j$ and $u_j$. If this is done,
then many terms cancel out and
we are left with
\bena
B v_i(x_1, x_2) &=& \sum_{j=1}^{i-1} \C_0(x_1, x_2) \Big( s_j s_{i-j} \otimes id + id \otimes
s_j s_{i-j} \Big) - s_{i-j}s_j \Big( \C_0(x_1, x_2) \Big) \non\\
&=& -bu_i(x_1, x_2) \, ,
\eena
which is what we wanted to show.
We finally prove the relation $Bw_i = -bv_i$.
Using the definition of $b$ and of $v_i$, we see after some
manipulations that $bv_i$ can be brought into the form
\bena
&&bv_i(x_1, x_2, x_3) = \\
&&\sum_{j=1}^{i-1} -bs_j(x_1, x_3) \Big( id \otimes \C_{i-j}(x_2,x_3) \Big)
                   +bs_j(x_2, x_3) \Big(\C_{i-j}(x_1, x_2) \otimes id \Big)                   \non\\
&&\sum_{j=1}^{i-1} +\C_j(x_2, x_3) \Big(bs_{i-j}(x_1, x_2) \otimes id \Big)
                   -\C_j(x_1, x_3) \Big(\gamma \otimes bs_{i-j}(x_2, x_3) \Big)                   \non\\
&&\sum_{j=1}^{i-1} -b\C_j(x_1, x_2, x_3) \Big(s_{i-j} \otimes id \otimes id +
       \gamma  \otimes s_{i-j} \otimes id +
       \gamma \otimes \gamma \otimes s_{i-j}\Big)                   \non\\
&&\sum_{j=1}^{i-1} s_j \Big( b\C_{i-j}(x_1, x_2, x_3) \Big)                   \,\, , \non
\eena
where $(x_1, x_2, x_3) \in \F_3$. On this domain may substitute
the assumption of the lemma that $bs_j + B\C_j = v_j$ and that $b\C_j = w_j$
for all $j<i$. This results in the equation
\bena
&&bv_i(x_1, x_2, x_3) = \\
&&\sum_{j=1}^{i-1} +B\C_j(x_1, x_3) \Big( id \otimes \C_{i-j}(x_2,x_3) \Big)
                   -B\C_j(x_2, x_3) \Big(\C_{i-j}(x_1, x_2) \otimes id \Big)                   \non\\
&&\sum_{j=1}^{i-1} -\C_j(x_2, x_3) \Big(B\C_{i-j}(x_1, x_2) \otimes id \Big)
                   +\C_j(x_1, x_3) \Big(\gamma \otimes B\C_{i-j}(x_2, x_3) \Big)                   \non\\
&&\sum_{j=1}^{i-1} -v_j(x_1, x_3) \Big( id \otimes \C_{i-j}(x_2,x_3) \Big)
                   +v_j(x_2, x_3) \Big(\C_{i-j}(x_1, x_2) \otimes id \Big)                   \non\\
&&\sum_{j=1}^{i-1} +\C_j(x_2, x_3) \Big(v_{i-j}(x_1, x_2) \otimes id \Big)
                   -\C_j(x_1, x_3) \Big(\gamma \otimes v_{i-j}(x_2, x_3) \Big)                   \non\\
&&\sum_{j=1}^{i-1} -w_j(x_1, x_2, x_3) \Big(s_{i-j} \otimes id \otimes id  +
          \gamma  \otimes s_{i-j} \otimes id + \gamma \otimes \gamma \otimes s_{i-j}\Big)                   \non\\
&&\sum_{j=1}^{i-1} +s_j \, w_{i-j}(x_1, x_2, x_3)                   \,\, . \non
\eena
We compute the first four terms in the expression on the right
hand side as
\bena
&=&\sum_{j=1}^{i-1} +s_0\C_j(x_1, x_3) \Big( id \otimes \C_{i-j}(x_2,x_3) \Big)                   \non\\
&&\sum_{j=1}^{i-1}  -\C_j(x_1, x_3) \Big(s_0 \otimes \C_{i-j}(x_1, x_2) \Big)                   \non\\
&&\sum_{j=1}^{i-1} -s_0\C_j(x_2, x_3) \Big(\C_{i-j}(x_1, x_2) \otimes id \Big)                   \non\\
&&\sum_{j=1}^{i-1} +\C_j(x_2, x_3) \Big(\gamma \, \C_{i-j}(x_1, x_2) \otimes s_0 \Big)                   \non\\
&&\sum_{j=1}^{i-1} +\C_j(x_2, x_3) \Big( \C_{i-j}(x_1,x_2)(s_0 \otimes id) \otimes id \Big)                   \non\\
&&\sum_{j=1}^{i-1} +\C_j(x_2, x_3) \Big(\C_{i-j}(x_1, x_2)(\gamma \otimes s_0) \otimes id \Big)                   \non\\
&&\sum_{j=1}^{i-1} -\C_j(x_1, x_3) \Big(\gamma \otimes \C_{i-j}(x_2, x_3) (s_0 \otimes id) \Big)                   \non\\
&&\sum_{j=1}^{i-1} -\C_j(x_1, x_3) \Big(\gamma \otimes \C_{i-j}(x_2, x_3) (\gamma \otimes s_0) \Big) = -Bw_i(x_1, x_2, x_3) \, .
\eena
The remaining terms cancel if we substitute the expressions $bs_j + B\C_j = v_j$ and $b\C_j = w_j$
for $v_j, w_j$ for $j<i$. Thus, we
have shown that $bv_i = -Bw_i$, and this concludes the proof of the lemma.
\qed

\section{Euclidean invariance}\label{euclidinvariance}

Above, we have defined quantum field theory by a collection of
OPE-coefficients subject to certain axiomatic requirements, and
we have pointed out that the essential information is contained in the
2-point coefficients $\C(x_1, x_2)$. The main condition that these
conditions have to satisfy is the associativity condition~\eqref{maincondition}.
They also have to satisfy the condition of Euclidean invariance. We will now
explain how that condition can be used to simplify the coefficients $\C(x_1, x_2)$,
and how to reformulate the associativity condition in terms of the simplified
coefficients.

Let us denote the components of $\C(x_1, x_2)$ in a basis of $V$ by
$C^c_{ab}(x_1, x_2)$. We use Euclidean invariance to write these 2-point OPE
coefficients as
\ben\label{cdecomp}
C_{ab}^c(x_i, x_j) = \sum_I
\left[
\begin{matrix}
\hat c \\
\hat a \,\,\,\, \hat b
\end{matrix}
; \,\,I
\right] (\hat x_{ij}) \cdot
f_{ab}^c(I; r_{ij}) \, .
\een
Here, the quantity in brackets is an invariant tensor
\ben\label{invarianttensor}
\left[
\begin{matrix}
i \\
j \,\,\,\, k
\end{matrix}
;\,\,
I
\right]:
S^{D-1} \mapsto V_i^{} \otimes V_j^{} \otimes V_k^* \, ,
\een
meaning that it satisfies the transformation law
\ben
\left[
\begin{matrix}
i \\
j \,\,\,\, k
\end{matrix}
;\,\,
I
\right] (g \hat x) = R_i^*(g) R_j^{}(g) R_k^{}(g)
\left[
\begin{matrix}
i \\
j \,\,\,\, k
\end{matrix}
;\,\,
I
\right] (\hat x) \, ,
\een
for all $\hat x \in S^{D-1}$, and all $g$ in the
covering (spin) group of $SO(D)$. The quantities $f_{ab}^c: \mr_+ \to \mc$ are analytic functions valued in
the complex numbers, $r_{ij}=|x_i - x_j|$, $\hat x_{ij} = x_{ij}/r_{ij}$, and
$I$ is an index that labels the space of invariant tensors
on the $(D-1)$-dimensional sphere.

In the following, we will restrict attention to
the case $D=3$ for pedagogical purposes, since
the representation theory of the corresponding
spin group $SU(2)$ is most familiar.
In the case $D=3$, the representation
labels may be identified with spins $\in \frac{1}{2} \mn$, and
the representation spaces are $V_j = \mc^{2j+1}$.
A basis of invariant tensors~\eqref{invarianttensor} is labeled
by a pair of spins
$I=[l_1 l_2] \in \frac{1}{2} \mn \times \frac{1}{2} \mn$,
and is given by
\ben\label{o3inv}
\left[
\begin{matrix}
j_1 \\
j_2 \,\,\,\, j_3
\end{matrix}
;\,\,I
\right]
(\hat x) =
\left\{
\begin{matrix}
l_1 \\
j_2 \,\,\,\, j_3
\end{matrix}
\right\}
\left\{
\begin{matrix}
j_1 \\
l_1 \,\,\,\, l_2
\end{matrix}
\right\} Y_{l_2}(\hat x)
\een
in terms of the Clebsch-Gordan coefficients ($3j$-symbols) of $SU(2)$ and the
spherical harmonics $Y_{lm}$ on $S^2$. Here
we have suppressed the magnetic quantum numbers, and as
everywhere in what follows,
magnetic quantum numbers associated with spins are summed
over if the spins appear twice. In the above example, the invariant
tensor should have 3 additional indices for the magnetic quantum
numbers associated with the representations
$j_1, j_2, j_3$, which have been suppressed.
The magnetic quantum numbers associated with $l_1, l_2$ are
contracted in the above expression, because each of these spins
appears twice.

\begin{figure}
\setlength{\unitlength}{1cm}
\begin{center}
\includegraphics[width=4.5in]{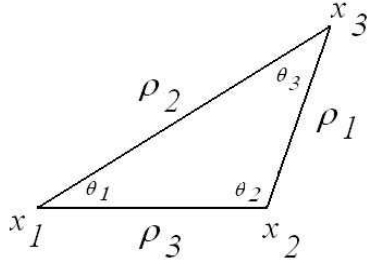}
\end{center}
\caption{The triangle spanned by $x_1, x_2, x_3$.}
\label{triangle}
\end{figure}

The decomposition~\eqref{cdecomp} provides a split of the 2-point
OPE coefficients into the purely representation theoretic tensor part
$[
\therefore \,\, ;\,\,I
]$
determined entirely by the representation theory of $SU(2)$,
and the dynamical part $f^c_{ab}$, which is a scalar
function that is holomorphic in the radial variable $r \in \mr_+$.
It is clear that it should be possible to formulate our associativity
condition in terms of these functions $f^c_{ab}$, as the
tensor coefficients are determined entirely in terms of group theory. To present
the resulting associativity conditions on $f^c_{ab}$
in a reasonably short form, we introduce the
notation $\rho_1= r_{23}, \rho_2 = r_{13}, \rho_3 = r_{12}$ for the
side lengths and
\ben
\theta_1 = \arccos \frac{\rho_2^2 + \rho_3^2 - \rho_1^2}{2 \rho_2 \rho_3}, \quad
{\rm etc.}
\een
for the angles of the triangle in $\mr^3$ spanned by $x_1, x_2, x_3$, see fig.~\ref{triangle}.

We also denote the spin associated with a field
$\phi_a$ by $\hat a \in \frac{1}{2} \mn$.
Then the associativity condition~\eqref{maincondition} is equivalent
to the following condition:
\bena\label{conscond3d}
&& \sum_b \sum_{j_1, j_2, j_5} \left\{
\begin{matrix}
j_6 & j_2 & \hat a_4\\
j_7 & j_5 & j_1
\end{matrix}
\right\} \left\{
\begin{matrix}
j_3 & j_5 & \hat b \\
j_1 & \hat a_3 & j_6
\end{matrix}
\right\} \,
{\rm P}_{j_1}(\cos \theta_2)\\
&& \times
f_{a_1 a_2}^{b}\Big(\rho_3; [j_3 j_1] \Big)
f_{b a_3}^{a_4}\Big(\rho_1; [j_5 j_2] \Big)=\non\\
&& \sum_b \sum_{j_1, j_2, j_4, j_5}
\left\{
\begin{matrix}
j_6 & j_2 & \hat a_4\\
j_7 & j_5 & j_1
\end{matrix}
\right\} \left\{
\begin{matrix}
j_4 & j_5 & \hat a_5 \\
j_1 & \hat a_2 & j_6
\end{matrix}
\right\}
\left\{
\begin{matrix}
\hat a_1 & j_6 & j_4 \\
\hat a_3 & \hat a_2 & j_5
\end{matrix}
\right\}
\,
{\rm P}_{j_1}(\cos \theta_3) \non\\
&&\times
f_{a_1 a_3}^{b}\Big(\rho_2; [j_4 j_1] \Big)
f_{a_2 b}^{a_4}\Big(\rho_1; [j_5 j_2] \Big) \non \, ,
\eena
in the domain $\rho_3 < \rho_1 < \rho_2$.
Here, the expressions in brackets denote the
well-known $6j$-symbols for $SU(2)$,
\ben
\left\{
\begin{matrix}
j_1 & j_2 & j_3\\
j_4 & j_5 & j_6
\end{matrix}
\right\}
=
\left\{
\begin{matrix}
j_3 \\
j_1 \,\,\,\, j_2
\end{matrix}
\right\}
\left\{
\begin{matrix}
j_4 \\
j_3 \,\,\,\, j_5
\end{matrix}
\right\}
\left\{
\begin{matrix}
j_5 \,\,\,\, j_2 \\
j_6
\end{matrix}
\right\}
\left\{
\begin{matrix}
j_6 \,\,\,\, j_1 \\
j_4
\end{matrix}
\right\} \, .
\een
The expressions
${\rm P}_j(z) = {}_2 F_1(-j, j+1, 1; (1-z)/2)$ are the Legendre polynomials.
A similar form of the associativity condition can be obtained for arbitrary
dimensions $D \ge 3$, the only essential difference being that we now encounter
the $6j$-symbols for the spin groups of $SO(D)$ for general $D$. The case $D=2$ is
an exceptional case and the corresponding expression is much simpler, owing
to the fact that the representation theory $SO(2)$ and its covering $\mr$ is
much simpler.

If we let $|a|$ be the dimension of the field $\phi_a$, then the scaling axiom for the
OPE-coefficients implies the relation
\ben
f_{ab}^c(r) = O(r^{|c|-|a|-|b|}) \, .
\een
In the case of the free quantum field theory in 3 dimensions defined by the
Lagrangian $L = \frac{1}{2}(\partial \varphi)^2$, the coefficients are in fact monomials and
are given by $f_{ab}^c(r) = \zeta_{ab}^c r^{|c|-|a|-|b|}$ for some complex constants
$\zeta^c_{ab}$, see section~\ref{freefield} for details. Furthermore, one can show that~\cite{Hollands06},
for the coefficients of the perturbatively defined theory with Lagrangian $L = \frac{1}{2} (\partial \varphi)^2 -
\frac{1}{6} \lambda \varphi^6$ and dimensionless $\lambda$, the coefficients take the form
\ben
f_{ab}^c(r) = p_{ab}^c(\log r, \lambda) r^{|c|-|a|-|b|} \, ,
\een
with $p_{ab}^c$ a polynomial in two variables whose degree is $n$ in $\lambda$ if we
compute the coefficients to $n$-th order in perturbation theory, and whose degree in
$\log r$ is no more than $n$ at $n$-th order. The associativity condition~\eqref{conscond3d}
is a quadratic constraint for these polynomials $p_{ab}^c$ at each
arbitrary but fixed order in perturbation theory.

If there are dimensionful parameters in the
lagrangian, those would effectively be treated as other perturbations in our framework.
For example, for the Lagrangian $L = \frac{1}{2} (\partial \varphi)^2 + \frac{1}{2} m^2 \varphi^2 +
\frac{1}{6} \lambda \varphi^6$, the coefficients take the form
\ben
f_{ab}^c(r) = p_{ab}^c(r, \log r, m^2, \lambda) r^{|c|-|a|-|b|} \, ,
\een
where $p_{ab}^c$ is again a polynomial in all four variables at $n$-th
perturbation order in $m^2$ and $\lambda$. Each term in this polynomial
containing a power $m^{2k}$ contains exactly a power of $r^{2k}$ so as
to make each term "dimensionless" (with the logarithms and $\lambda$ not counting
as having a dimension).

\section{The fundamental left (vertex algebra) representation}\label{leftrep}

In the previous sections, we have elaborated on our definition of
quantum field theory in terms of consistency conditions. Our formulation involved
only the OPE coefficients such as $C_{ab}^c$. To motivate our constructions, we
sometimes wrote formal relations like
\ben
\text{``$\phi_a(x_1) \phi_b(x_2) = \sum_c C_{ab}^c(x_1, x_2) \, \phi_c(x_2)$''} \quad .
\een
But these relations were only heuristic, in the sense that none of our proposed
properties of the OPE coefficients relied on the existence or properties of the
hypothetical operators $\phi_a$, which were only "dummy variables".
As we have emphasized, our approach is similar
to the standard viewpoint taken in algebra that an abstract algebra $\A$ is entirely defined in
terms of its product---i.e., a linear map $m: \A \otimes \A \to \A$ subject to
the associativity condition. But, as in our case, the algebra elements
need not be represented a priori by linear operators on a vector space. Representations in
the context of an algebra are an additional structure defined as
linear maps $\pi: \A \to {\rm End}(H)$ from the algebra to
the linear operators on a vector space $H$, subject to the condition $\pi[m(A,B)] = \pi(A)\pi(B)$. It is natural
to ask whether there is a construction similar to a representation also in our context.
We shall show in this section that there is indeed a certain "canonical" construction, which
has some features in common with an algebra representation, and which will be useful
in the next section. We will refer to this construction as the "fundamental left-" or
"vertex algebra representation".

\begin{defn}
Let $|v\rangle \in V$ be an arbitrary vector. We define a corresponding {\em vertex operator}
$\V(x, v): V \to V$ by the formula
\ben
\V(x, v)|w\rangle = \C(x, 0)(|v\rangle \otimes |w\rangle) \, ,
\een
for all $x \neq 0$. In a basis $\{ |v_a\rangle \}$, the matrix representing the vertex operator is hence given by
\ben
[\V(x, v_a)]_b^c := C_{ab}^c(x,0) \,\,\,\,\,.
\een
This is our {\it fundamental left-} or {\it vertex algebra representation}.
\end{defn}

Using the consistency condition~\eqref{maincondition},
one can immediately show that
\ben\label{lalblc1}
\V(x, v_a) \V(y, v_b) = \sum_c C_{ab}^c(x,y) \, \V(y, v_c) \, ,
\een
for $0<|x-y|<|y|<|x|$, or equivalently that
\ben\label{lalblc}
\V(x, v_a) \V(y, v_b) = \V(y, \V(x-y, v_a)v_b ) \, .
\een
Thus, by eq.~\eqref{lalblc1}, the vertex operators operators $\V(x, v_a): V \to V$ satisfy the operator product expansion.
The fact that the OPE coefficients in this expansion are precisely the matrix elements of the vertex operators
themselves is expressed in the second relation~\eqref{lalblc}. This quadratic relation is the key
axiom in the theory of vertex operator algebras, see \cite{vertex1, vertex2, vertex3, vertex4}.
Because of eq.~\eqref{lalblc1}, we may formally view the vertex operators as forming a "representation" of the heuristic field
operators, i.e., formally "$\pi(\phi_a(x)) = \V(x, v_a)$" is a "representation" of the "algebra"
defined by the OPE coefficients. This "representation" is in some sense analogous to the
GNS-representation~(see e.g.~\cite{Haag}) for $C^*$-algebras. However, we emphasize that
in our case, $V$ is not in a natural way a Hilbert space, and should not be confused with the
physical Hilbert space obtained via the Osterwalder-Schrader reconstruction theorem, see our
remarks in section~\ref{axiomatic}. We will further develop the analogy of our approach to
the theory of vertex operator algebras in a forthcoming paper~\cite{Olbermann}.

\section{Example: The free field}\label{freefield}
Let us now explain our approach to quantum field theory in
a simple example, namely that of a free hermitian bosonic scalar field in $D$ dimensions
classically described by the field equation
$$\square \varphi = 0,$$
with $\square = \delta^{\mu\nu} \partial_\mu \partial_\nu$.
The aim is to present explicitly the OPE coefficients $\C(x_1, x_2)$
for this model.
This section is joint work with H. Olbermann and details will appear elsewhere.
We begin by describing the space $V$ of fields in our case, assuming $D>2$ for simplicity. The
case $D=2$ can be treated analogously, with only minor modifications.

\begin{defn}
$V$ is the defined to be the commutative, unital, $\mc$-module
generated as a module (i.e., under addition, multiplication and scalar multiplication)
by formal expressions of the form $\partial_{\{\mu_1} \dots \partial_{\mu_N\}} \varphi$,
and unit $\myid$, where $\mu_i = 1, \dots, D$ and a curly bracket denotes the
totally symmetric, trace-free part, i.e. by definition,
\ben
\delta^{\mu_i \mu_j}\,  \partial_{\{\mu_1} \dots \partial_{\mu_N\}} \varphi = 0 \, .
\een
\end{defn}

\noindent
The trace free condition has been imposed because any trace would give rise
to an expression containing $\square \varphi$, which we want to vanish in
order to satisfy the field equation on the level of $V$.
A basis of $V$ as a $\mc$-vector space can e.g. be given as follows. First,
let us choose a basis of totally symmetric, trace-free, rank-$l$ tensors
in $\mr^D$ for any $l \ge 0$. For a given $l \ge 0$, this space has
dimension $N(l,D)$, where
\ben
N(l,D) =
\begin{cases}
1 & \text{for $l=0$}\\
\frac{(2l+D-2)(l+D-3)!}{(D-2)!l!} & \text{for $l>0$.}
\end{cases}
\een
We denote the basis elements by $t_{l,m}, m=1, \dots, N(l,D)$, and
we assume for convenience that they are orthonormal with respect to the natural hermitian
inner product on $(\mr^{D})^{\otimes l}$ coming from the Euclidean metric
on $\mr^D$, i.e. $\bar t_{l',m'} \cdot t_{l,m} = \delta_{ll'} \delta_{mm'}$.
A basis of $V$ is then given by $\myid$, together with the elements
\ben\label{vadef}
|v_a \rangle =  \prod_{l,m} (a_{l,m}!)^{-1/2}
\left( c_l^{-1/2} \, t_{l,m} \cdot \partial^l \varphi \right)^{a_{l,m}} \, \quad \, ,
\een
where $a = \{ a_{l,m} \mid l \ge 0, 0 < m \le N(l,D) \}$ is a multi-index of
non-negative integers, only finitely many of which are non-zero. For later
convenience, we also set
\ben
c_l = \frac{2^l \, \Gamma(l+1) \Gamma(l+D/2-1)}{\Gamma(D/2-1)}  \, .
\een
The canonical dimension of $|v_a \rangle$ is defined as
\ben
|a| = \sum_{l,m} a_{l,m}[(D-2)/2+l] \, .
\een
It is possible to formally view $V$ as a "Fock-space", with $a_{l,m}$ the "occupation numbers"
of the "mode" labeled by $l,m$. On this Fock-space, one can then define
creation and annihilation operators $\a_{l,m} , \a_{l,m}^+: V \to V$ as usual.
These are defined explicitly by
\bena
\a_{l,m} |v_a \rangle &:=& (a_{l,m})^{1/2}
\, |v_{a-e_{l,m}} \rangle \\
\a_{l,m}^+ |v_a \rangle   &:=& (a_{l,m}+1)^{1/2} \, |v_{a+e_{l,m}} \rangle
\eena
where $e_{l,m}$ is the multiindex with a unit entry at position $l,m$ and zeros elsewhere.
They satisfy the standard commutation relations
\ben
\left[ \a_{l,m}^{}, \a_{l',m'}^{+} \right] =
\delta_{ll'}\delta_{mm'} \,\, id \, , \quad
\left[ \a_{l,m}^+, \a_{l',m'}^+ \right] =
\left[ \a_{l,m}^{}, \a_{l',m'}^{} \right] = 0
\een
where $id$ is the identity operator on $V$. The ``vacuum'' vector $|0\rangle$ in this Fock space
by definition corresponds to the identity operator $\myid \in V$.

To present the OPE coefficients of the model, it is further convenient to
introduce spherical harmonics in $D$ dimensions. The most straightforward way to do this is
as follows. Let $l \in \mn_0$, and let $h_l(x) \in \mc[x]$ be a harmonic polynomial on $\mr^D$ that
is homogeneous of degree $l$, meaning that $\square h_l(x) = 0$, and that $h(\lambda x) =
\lambda^l h_l(x)$ for all $\lambda \in \mr_+$. It is not difficult to see that the vector
space spanned by such polynomials is of dimension $N(l,D)$. We let $h_{l,m}(x), 0<m\le N(l,D)$ be
a basis of this vector space
and we define the (scalar) spherical harmonics $Y_{l,m}:S^{D-1} \to \mc$ to be the restriction
of the corresponding harmonic polynomials to the $(D-1)$-dimensional sphere. We normalize
the spherical harmonics to turn them into an orthonormal basis on the sphere, in the natural
$L^2$-inner product. The spherical harmonics are closely related to the trace free
symmetric tensors $t_{l,m}$ in $(\mr^D)^{\otimes l}$ that were introduced above. In fact,
we may choose
\ben\label{ylmdef}
Y_{l,m}(\hat x) =
k_l \, \bar t_{l,m} \cdot \hat x^{\otimes l} \, ,
\een
for some normalization constant $k_l$. With this notation in
place, we now explicitly present the OPE coefficients $\C(x_1, x_2)$ for this model. For this, it is
sufficient to present the vertex operators (left-representatives) $\V(x, v_a):V \to V$ for all $|v_a\rangle \in V$, since the
matrix elements $[\V(x, v_a)]_b^c = C_{ab}^c(x,0)$ are by definition just the OPE coefficient components, see
sec.~\ref{leftrep}.
First, we give the formula for $\V(x, \varphi)$ corresponding to the basic field $\varphi \in V$. This is defined by
\begin{multline}
\V(x, \varphi) = \sqrt{{\rm vol}(S^{D-1})} \,
 \sum_{l=0}^{\infty} \, \sum_{m=1}^{N(l,D)} \sqrt{\frac{D-2}{2l+D-2}} \times \\
\Big[ r^{l} Y_{l, m}(\hat x) \, \a_{l,m}^{+} + r^{-l-D+2} \overline{ Y_{l, m}(\hat x) } \, \a_{l,m}^{} \Big] \, .
\end{multline}
We will "derive" this formula
from the standard quantum field theory formalism in a future paper~\cite{Olbermann}.
Accidentally, this has precisely the familiar form for a free field operator, with
an "emissive" and an "absorptive" piece, which should not come as a surprise,
since $\V(x, \varphi)$ is in a sense the "representative"
of the (formal) field operator $\varphi(x)$ on $V$. Actually, if we furthermore write $r = \e^t$, then this is
precisely the formula for a free field operator on the manifold $\mr \times S^{D-1}$ with
"time" $t$ formally imaginary. We will pursue this analogy elsewhere.

For a general element in $V$, we now give a corresponding formula for the
vertex operator. It is defined by $\V(x, \myid) = id$ for the identity element, and by
\ben\label{Ladeffree}
\V\Big(x, \prod_i \partial^{l_i} \varphi  \Big) =
\,\,
: \prod_{i}
\partial^{l_i} \V\Big(x, \varphi \Big) :
\,\,
 \,  .
\een
for a general field monomial.
Here, the following notation is used. The double dots $: \dots :$ mean
"normal ordering", i.e., all creation operators are to the right of all annihilation operators.
Again, one can derive this formula using the standard quantum field theory formalism. The
OPE coefficients for the free field are consequently given by $C_{ab}^c(x_1, x_2) := [\V(x_1-x_2, v_a)]_b^c
= \langle v^c | \V(x_1-x_2, v_a) | v_b \rangle$ or more explicitly by
\ben\label{cabcfreedef}
C_{ab}^c(x_1, x_2) := \Big\langle 0 \Big|
\prod_{l,m} (\a_{l,m}^{})^{c_{l,m}}
\,
\V(x_1-x_2, v_a)
\,
\prod_{l,m} (\a_{l,m}^+)^{b_{l,m}}
\Big| 0 \Big\rangle \, .
\een
We now state that the so-defined OPE-coefficients  satisfy
our consistency condition:
\begin{thm}
Let $\V(x, v): V \to V$ be defined for our model by formula~\eqref{Ladeffree},
and let the OPE-coefficients $C_{ab}^c(x_1, x_2)$ be defined by eq.~\eqref{cabcfreedef}.
Then the OPE coefficients satisfy the consistency condition~\eqref{maincondition}.
Equivalently, the vertex algebra condition~\eqref{lalblc} holds for the free
field vertex operators $\V(x, v_a)$.
\end{thm}

\noindent
{\em Proof:} The proof of this theorem is essentially a longish but straightforward computation, using
various standard identities for the $D$-dimensional spherical harmonics. We will give a
complete proof in \cite{Olbermann}.

\section{Interacting fields}\label{interactingfields}

In the previous section, we have presented the (2-point) OPE coefficients in the example of
a free quantum field associated with the classical equation $\square \varphi = 0$.
It is clearly of interest to know what would be the corresponding coefficients for
a field associated with a non-linear equation such as
\ben
\square \varphi = \lambda \varphi^p
\een
where $p$ is some non-negative integer. As has been appreciated for a long time, the construction of
a quantum field theory (and hence in particular of the OPE) associated with such
an equation is extremely difficult, and has only been accomplished so far for
certain values of $p,D$ where the theory has a particularly simple behavior. However, one can treat
$\lambda$ as a formal perturbation parameter, and try to construct the OPE coefficients
in the sense of formal power series in $\lambda$ as we have outlined in general terms in section~\ref{hochschild}.
Here we would like to outline how a field equation can help to actually determine
the formal power series in the theory described by a field equation of the above type.
Some of the ideas in this section go back, in preliminary form, to discussions with
N.~Nikolov, and also to joint work with H.~Olbermann, which will be published in~\cite{Olbermann}.

As we have seen in section~\ref{leftrep}, the 2-point
OPE coefficients $\C(x_1, x_2)$ contain the same information as the
corresponding vertex operators $\V(x, v)$. In perturbation theory, they
are given by formal power series
\ben
\V(x, v) = \sum_{i=0}^\infty \V_i(x, v) \, \lambda^i \, ,
\een
where each $\V_i(x, v)$ is a linear map $V \to V$, and where
$\V_0(x, v)$ is given by the free field vertex operator
defined in the previous section~\ref{freefield}. As discussed in
subsection~\ref{subfieldeq}, we expect
that the field equation implies:
\ben\label{fieldeqvertex}
\V_i(x, \varphi) = \square^{-1} \V_{i-1}(x, \varphi^p ) \, .
\een
More precisely, in this section we {\em assume} the existence of
$\V_i$ satisfying this equation, and we also {\em assume} that the
consistency condition~\eqref{lalblc} is satisfied order-by-order;
in vertex operator notation
\ben
\sum_{j=0}^i \V_j( y, v_a) \V_{i-j}(x, v_b)
=
\sum_{j=0}^i \V_{i-j}\Big(x, \V_j ( y-x, v_a  ) v_b  \Big) \, .
\een
As we will now show, these assumptions will allow us to inductively
determine the actual form of the vertex operators order by
order in $i$. But before we do this, we must explain a point
related to the choice of $V$ in for our interacting theory.
Recall that, in the underlying free theory with $\lambda=0$, $V$ was spanned by formal monomials in
$\varphi$ and its derivatives $\partial_{\{\mu_1} \dots \partial_{\mu_N\}} \varphi$,
where he curly brackets denote the trace-free
part of a tensor. In the free theory, we considered the trace free part only,
since any trace gives rise to a factor of $\square \varphi$ in such a monomial, $v$,
and the corresponding vertex operator $\V_0(x,v)$ then vanishes (essentially by definition). However,
for the interacting theory, we must be more careful and allow also traces, i.e.,
we also consider vertex operators whose arguments are formal monomials in $\varphi$ and
its derivatives $\partial_{\mu_1} \dots \partial_{\mu_N} \varphi$.
This enlarged space of objects, $\widehat V$, is a commutative unital differential
module (with derivations $\partial_\mu, \mu=1, \dots, D$ acting in the usual way), and
the vertex operators $\V_i(x, v)$ should now be considered as linear maps $\widehat V \owns v \mapsto
\V_i(x, v) \in {\rm End}(\widehat V)$. We then also {\em assume} to have a relation
\ben
\partial_\mu \, \V_i(x, v) = \V_i(x, \partial_\mu v) \quad, \quad \mu=1, \dots, D \, ,
\een
where the symbol $\partial_\mu$ denotes a genuine partial $x$-derivative on the left side,
while it is the derivation on the differential module $\widehat V$ on the right side. For details, we
refer to~\cite{Olbermann}. To lighten the notation, we will drop the caret on $\widehat V$ again
for the remaining part of the section.

To make sense of eq.~\eqref{fieldeqvertex}, we first of all need to define the inverse of the Laplace operator.
We rewrite it in $D$-dimensional polar coordinates, and
we furthermore assume that we can expand each vertex operator in spherical harmonics and
coefficients in the ring $\mc[r, 1/r, \log r] \otimes {\rm End}(V)$.
Then the vertex operators schematically take the form
\ben
\V_i(x, v) = \sum A_{i,l,m,j,k}(v) r^k (\log r)^j Y_{l,m}(\hat x) \, ,
\een
with $A_{i,l,m,j,k}(v) \in {\rm End}(V)$.
We define the action of the inverse Laplacian on such expressions by putting\footnote{
It follows from the inductive construction that, if we take any matrix element 
of $\V_i$ between $\langle v^a|$ and $|v_b \rangle$, then there remain only finitely 
many terms in the above sum. Hence, we may take the inverse of the Laplacian term-by-term
without problem. 
}
\bena
&&\square^{-1} [r^k (\log r)^j Y_{l,m}(\hat x) ] :=
j! Y_{l,m}(\hat x) \times \non\\
&& \times
\begin{cases}
(-1)^{j+1} r^l \sum_{i=0}^{j+1} \frac{(-1)^i \log^i r}{i!(2l+D-2)^{j-i+2}}  &
\text{if $k=l-2$}\\
-r^{-l-D+2} \sum_{i=0}^{j+1} 
\frac{\log^i r}{i!(2l+D-2)^{j-i+2}} &
\text{if $k=-l-D$}\\
r^{k+2} \sum_{i=0}^j \sum_{n=0}^i \frac{(-1)^{i-n} \log^{j-i} r}{(j-i)!(l-k-2)^{n+1}(l+k+D)^{i-n+1}} &
\text{otherwise.}
\end{cases}
\eena
This is a left inverse  for the Laplacian. Any other left
inverse can differ from this one only by terms in the kernel of
$\square$, i.e. a harmonic polynomial of $x$ with values in
${\rm End}(V)$.

Let us now assume inductively that we have constructed all the
vertex operators $\V_j(x, v)$ up to
order $j=i-1$. The vertex operator $\V_i(x, \varphi)$ is then given
by eq.~\eqref{fieldeqvertex}. Next, we would like to determine all other vertex operators $\V_i(x, v )$,
where $|v \rangle \in V$ is a general element. For this, we perform, at fixed $i$,
an induction in the dimension $\Delta(v)$. Thus, let us assume that we
have succeeded in constructing all vertex operators up to dimension $d$, and let us assume
for the sake of concreteness that we are in $D=4$, so that $\Delta(\varphi) = 1$.
We may hence assume that $d \ge 2$. We may write a general field of dimension $d+1$
as a linear combination of fields of the form $v = w \partial^l \varphi$,
or of the form $v = \partial^{l+1} w$. In both cases,
$w$ has dimension $d-l$, and so $\V_j(x, w)$ is inductively
known for $0 \le j \le i$. In the second case, we must have
$\V_i(x, v) = \partial^{l+1} \V_i(x, w)$. In the first case, the
consistency condition gives
\ben
\sum_{j=0}^i \V_j\Big( y, \partial^l \varphi  \Big) \V_{i-j}\Big(x, w \Big)
=
\sum_{j=0}^i \V_{i-j}\Big(x, \V_j ( y-x, \partial^l \varphi  ) w  \Big) \, .
\een
By the inductive hypothesis, all operators on the left side of the equation are already known.
Now we investigate which operators are not already known on the right side. Evidently,
if $j \neq 0$, then all terms in the corresponding expression are known. If $j = 0$,
we look at the terms that survive in the limit $y \to x$. Using the definition of the
zeroth order vertex operators (free theory), we see that
\ben
\V_0 ( y-x, \partial^l \varphi  ) w  = w \partial^l \varphi  + \dots ,
\een
where the dots stand for the following terms: (a) terms that vanish as $|x-y| \to 0$ and
(b) a finite Laurent series in $1/|x-y|$ with coefficients that are vectors of dimension
$ \le d$. Let $P_d^j: V \to V$ denote the map which is the identity for
$j \neq 0$, which is the projector onto the subspace of vectors of dimension
$\le d$ for $j=0$. Then we can write:
\begin{multline}
\V_i(x, v) =
\lim_{y \to x} \Bigg[
\sum_{j=0}^i \V_j\Big( y, \partial^l \varphi  \Big) \V_{i-j}\Big(x, w  \Big) \\
-
\sum_{j=0}^i \V_{i-j}\Big(x, P_d^j \circ \V_j ( y- x, \partial^l \varphi ) w  \Big)
\Bigg] \, .
\end{multline}
Now all the terms on the right side are known inductively. We can hence determine all vertex operators
at order $i$, and hence to arbitrary orders.
This shows how we may construct inductively
the terms in the perturbation series starting from those of the free theory.

\section{Conclusions and outlook}

In this paper, we have suggested a new approach to general, non-conformal, quantum
field theories in terms of consistency conditions. These consistency conditions are
formulated in terms of the operator product expansion (OPE). We showed that these
conditions are quite powerful. For example, they can be used to characterize the possible perturbations of the
quantum field theory, and give rise to an efficient algorithm for explicitly
computing these coefficients.

This paper is just the beginning of a longer programme. In the future, we would like to
extend the ideas of the paper. In particular, it would be interesting to consider the
following issues:
\begin{itemize}
\item Generalization of our approach to curved space-time;
\item Convergence/Borel summability of the perturbation series;
\item Explicit perturbative calculations;
\item Incorporation of the renormalization group into our approach;
\item (Super-)conformal quantum field theories;
\item Perturbations of 2-dimensional conformal quantum field theories.
\end{itemize}
We intend to study these topics in future publications.
\\
\\
\\
\\
\\
\noindent
{\bf Acknowledgements:} I would like to thank N.~Nikolov for
extensive discussions on various topics in this paper. I would also like to
thank K.-H.~Rehren and R.~M.~Wald for discussions. I would especially
like to thank C.~Brouder for his careful reading of the manuscript, and
in particular for pointing out several sign errors in the first version.

\end{document}